\documentclass[aps,prb,twocolumn,superscriptaddress]{revtex4-1}
\usepackage{amssymb, amsmath, bm}
\usepackage{graphicx,onlyamsmath}

\usepackage{natbib}
\usepackage{xcolor}
\usepackage[normalem]{ulem}  

\begin{document}

\title{Disorder-induced topological phase transition in HgCdTe crystals}

\author{Sergey S.~Krishtopenko}
\affiliation{Laboratoire Charles Coulomb (L2C), UMR 5221 CNRS-Universit\'{e} de Montpellier, F- 34095 Montpellier, France}

\author{Mauro Antezza}
\affiliation{Laboratoire Charles Coulomb (L2C), UMR 5221 CNRS-Universit\'{e} de Montpellier, F- 34095 Montpellier, France}
\affiliation{Institut Universitaire de France, 1 rue Descartes, F-75231 Paris Cedex 05, France}

\author{Fr\'{e}d\'{e}ric Teppe}
\email[]{frederic.teppe@umontpellier.fr}
\affiliation{Laboratoire Charles Coulomb (L2C), UMR 5221 CNRS-Universit\'{e} de Montpellier, F- 34095 Montpellier, France}

\begin{abstract}

Using the self-consistent Born approximation, we study a topological phase transition appearing
in bulk HgCdTe crystals induced \emph{uncorrelated} disorder due to both randomly distributed impurities and fluctuations in Cd composition. By following the density-of-states evolution, we clearly demonstrate the topological phase transition, which can be understood in terms of the disorder-renormalized mass of Kane fermions. We find that the presence of heavy-hole band in HgCdTe crystals leads to the topological phase transition at much lower disorder strength than it is expected for conventional 3D topological insulators. Our theoretical results can be also applied to other narrow-gap zinc-blende semiconductors such as InAs, InSb and their ternary alloys InAsSb.
\end{abstract}

\maketitle

\section{\label{Sec:Int} Introduction}
Band structure of narrow-gap HgCdTe semiconductors in the vicinity of the $\Gamma$ point of the Brillouin zone is represented by specific type of three-dimensional (3D) fermionic excitations named Kane fermions~\cite{Int1,Int2,Int2b,Int3}. Although their band dispersion is closely resembling dispersion of Dirac fermions with an additional band, Kane fermions in the three-band approximation are in fact hybrids of pseudo-spin-1 and -1/2 Dirac fermions~\cite{Int4,Int5}. This results that unlike the pseudo-spin-1 Dirac fermions, Kane fermions are not protected by symmetry or topology, and their band-gap can be set at will by varying Cd concentration or external parameters such as temperature~\cite{Int2} and hydrostatic pressure~\cite{Int6,th1}. The band-gap vanishing represents the critical state corresponding to the topological phase transition between a trivial semiconductor and topological semimetal~\cite{Int7,Int8}.

The crucial influence of disorder on topological phase transitions was first discovered in two-dimensional (2D) systems hosting quantum spin Hall insulator state, like HgTe/CdHgTe quantum wells (QWs)~\cite{Int9}. Particularly, Li~\emph{et~al.}~\cite{Int9} have found that disorder may induce a topological phase transition if the QW is initially in the trivial semiconductor state. Then, Groth~\emph{et~al.}~\cite{Int10} have shown that the disorder-induced phase transition HgTe/CdHgTe QWs is caused by the quadratic terms $\propto k^2\sigma_z$ in 2D Bernevig-Hughes-Zhang (BHZ) Hamiltonian~\cite{Int11}. The latter is nothing than 2D Dirac Hamiltonian with additional quadratic corrections, required for the proper characterization of the topological state~\cite{Int12}. Later, disorder-induced phase transitions have also been studied in 3D topological insulators~\cite{TAI3Da,TAI3Db,TAI3Dc}, Dirac/Weyl semimetals~\cite{TAI3Dd,TAI3De,TAI3Df,TAI3Dg} and amorphous solids~\cite{TAI3Dh}. In the latter case, it has been demonstrated that an ``unconventional'' transition from a trivial insulator state to a topological semimetal, could be driven by disorder. Interestly, the discovered transitions correspond to weak-disorder topological transition, and thus can be treated within the self-consistent Born approximation (SCBA)~\cite{Int10,TAI3Dg,TAI3Dh,Int13,Int14,Int15}.

As clear from the above, the presence of disorder-induced topological phase transition is now evident for the systems, in which low-energy fermionic excitations are represented by Dirac or Weyl fermions. A question naturally arises whether such transition can occur in the system if the band structure is not represented by pseudo-spin-1/2 fermionic excitations. Below, considering the case of bulk HgCdTe crystals, we show the answer to the above question to be affirmative.

By using the SCBA and three-band Kane Hamiltonian in the \emph{continuous} representation, we directly calculate the density-of-states (DOS) as a function of the strength of \emph{uncorrelated} disorder in HgCdTe crystals. By following the DOS evolution, we directly demonstrate the band-gap vanishing with increasing of the disorder strength. We show that similar to the idea of Groth~\emph{et~al.}~\cite{Int10}, the topological phase transition in HgCdTe crystals can be also understood in terms of the disorder-renormalized mass of Kane fermions. Surprisingly, the presence of heavy-hole band in HgCdTe crystals leads to a phase transition at much lower disorder strength than it is expected for conventional 3D topological insulators~\cite{TAI3Da,TAI3Db,TAI3Dc}. As a source of the disorder, we consider the electrostatic potential of randomly distributed impurities, as well as fluctuations in the Cd composition~\cite{Int22,Int23} resulting in the band-gap fluctuations. Our theoretical results can be also applied to other narrow-gap zinc-blende semiconductors such as InAs, InSb and their ternary alloys InAsSb~\cite{Int24,Int25,Int26}.

\section{\label{Sec:Theory} Theory}
The band structure of narrow-gap zinc-blende semiconductors (including HgCdTe crystals) in the vicinity of the $\Gamma$ point of the Brillouin zone is qualitatively described by the three-band Kane Hamiltonian~\cite{th1}, which directly takes into account the $\Gamma_6$ and $\Gamma_8$ bands. In the basis set of Bloch amplitudes in the sequence $|\Gamma_6,+1/2\rangle$, $|\Gamma_6,-1/2\rangle$, $|\Gamma_8,+3/2\rangle$, $|\Gamma_8,+1/2\rangle$, $|\Gamma_8,-1/2\rangle$, $|\Gamma_8,-3/2\rangle$, the Hamiltonian is written as:
\begin{equation}
\label{eq:A1}
H_{3D}(\mathbf{k})=\begin{pmatrix}
H_{cc} & H_{cv} \\H_{cv}^{\dag} & H_{vv}\end{pmatrix},
\end{equation}
where the blocks $H_{cc}$ and $H_{vv}$ represent the contribution from the $\Gamma_6$ and $\Gamma_8$ bands, respectively, and the block $H_{cv}$ and the block $H_{cv}$ describes the band mixing. The block $H_{cc}$ is given by
\begin{equation}
\label{eq:A2}
H_{cc}=\left[E_c+\dfrac{\hbar^{2}}{2m_0}\mathbf{k}\left(2F+1\right)\mathbf{k}\right]I_{2{\times}2},
\end{equation}
where $I_{2{\times}2}$ is the $2{\times}2$ identity matrix, $E_c$ is the energy of conduction band edge conduction band profile, $\mathbf{k}=(k_x,k_y,k_z)$ and $F$ is a parameter accounting for contribution from remote bands. The block $H_{cv}$ has the form
\begin{equation}
\label{eq:A3}
H_{cv}=P\begin{pmatrix}
-\dfrac{\sqrt{2}k_{+}}{2} & \dfrac{\sqrt{6}k_{z}}{3} & \dfrac{\sqrt{6}k_{-}}{6} & 0 \\[6pt]
0 & -\dfrac{\sqrt{6}k_{+}}{6} & \dfrac{\sqrt{6}k_{z}}{3} & \dfrac{\sqrt{2}k_{-}}{2} \end{pmatrix},
\end{equation}
where $P$ is the Kane matrix element, $k_{\pm}=k_x{\pm}ik_y$. The block $H_{vv}$ is given by
\begin{equation}
\label{eq:A4}
H_{vv}=E_vI_{4{\times}4}+H^{(i)}_L,
\end{equation}
where $I_{4{\times}4}$ is the $4{\times}4$ identity matrix, $E_v$ is the energy of valence band edge and $H^{(i)}_L$ is the isotropic part of the Luttinger Hamiltonian~\cite{th2},
\begin{equation}
\label{eq:A5}
H^{(i)}_L=\dfrac{\hbar^2}{2m_0}\left[-\mathbf{k}\left(\gamma_1+\dfrac{5}{2}\gamma_2\right)\mathbf{k}+ 2(\mathbf{J}\cdot\mathbf{k})\gamma_2(\mathbf{J}\cdot\mathbf{k})\right],
\end{equation}
where $\gamma_1$, $\gamma_2$, and $\gamma_3$ are contributions to the Luttinger parameters from remote bands, and $\textbf{J}=(J_x,J_y,J_z)$ is the vector composed of the matrices of the angular momentum $3/2$. For simplicity, we neglect the small terms breaking rotational symmetry of $H_{3D}(\mathbf{k})$ in Eq.~(\ref{eq:A1}), as well as the terms resulting from the bulk inversion asymmetry of the unit cell of zinc-blende semiconductors~\cite{th3}.

Due to full rotational symmetry of $H_{3D}(\mathbf{k})$, its wave-function can be presented in the form:
\begin{equation}
\label{eq:A6}
\Psi_{3D}(\mathbf{k})=U_{z}(\phi)U_{y}(\theta)\Psi_{3D}(k),
\end{equation}
where
\begin{eqnarray}
\label{eq:A7}
U_{z}(\phi)=\begin{pmatrix}
\exp\left(-i\sigma_z\phi/2\right) & 0 \\0 & \exp(-i{J}_z\phi)
\end{pmatrix},\notag\\
U_{y}(\theta)=\begin{pmatrix}
\exp(-i\sigma_y\theta/2) & 0 \\0 & \exp(-i{J}_y\theta)
\end{pmatrix},
\end{eqnarray}
Here, $k=|\textbf{k}|$, $k_x=k\sin{\theta}\cos{\phi}$, $k_y=k\sin{\theta}\sin{\phi}$, $k_z=k\cos{\theta}$, while $\sigma_x$ and $\sigma_z$ represent the Pauli matrices.

The form of the wave function $\Psi_{3D}(\mathbf{k})$ in Eq.~(\ref{eq:A6}) allows one to introduce a new Hamiltonian $\tilde{H}_{3D}(k)$, depending only on $k$, using a unitary transformation $\tilde{H}_{3D}(k)=U_{y}(-\theta)U_{z}(-\phi)H_{3D}(\mathbf{k})U_{z}(\phi)U_{y}(\theta)$. In explicit form, the Hamiltonian $\tilde{H}_{3D}(k)$ is written as
\begin{widetext}
\begin{equation}
\label{eq:A8}
\tilde{H}_{3D}(k)=CI_{6{\times}6}+MI_{M}+\begin{pmatrix}
-(D+B)k^2 & 0 & 0 & Ak & 0 & 0 \\
0 & -(D+B)k^2 & 0 & 0 & Ak & 0 \\
0 & 0 & -\dfrac{\hbar^2k^2}{2m_{hh}} & 0 & 0 & 0 \\
Ak & 0 & 0 & -(D-B)k^2 & 0 & 0 \\
0 & Ak & 0 & 0 & -(D-B)k^2 & 0 \\
0 & 0 & 0 & 0 & 0 & -\dfrac{\hbar^2k^2}{2m_{hh}}
\end{pmatrix},
\end{equation}
\end{widetext}
where $I_{6{\times}6}$ is the $6{\times}6$, $I_M=\mathrm{diag}\{1,1,-1,-1,-1,-1\}$ is a diagonal matrix; other parameters are defined as $A=\sqrt{6}P/3$, $C=(E_c+E_v)/2$, $M=(E_c-E_v)/2$,
\begin{eqnarray}
\label{eq:A9}
D=\dfrac{\hbar^2}{2m_0}\dfrac{\gamma_{1}+2\gamma_{2}-2F-1}{2},~~\notag\\
B=-\dfrac{\hbar^2}{2m_0}\dfrac{\gamma_{1}+2\gamma_{2}+2F+1}{2},\notag\\
\dfrac{1}{m_{hh}}=\dfrac{1}{m_0}\left(\gamma_{1}-2\gamma_{2}\right).~~~~~~
\end{eqnarray}
The band parameters used in $\tilde{H}_{3D}(k)$ for Hg$_{1-x}$Cd$_{x}$Te crystals at different $x$ are summarized in Table~\ref{tab:1}. The Hamiltonian $\tilde{H}_{3D}(k)$ has three doubly-degenerate eigenvalues:
\begin{eqnarray}
\label{eq:A10}
E_{c,lh}=C-Dk^2
\pm\sqrt{\left(M-Bk^2\right)^2+A^2k^2},\notag\\
E_{hh}=C-M-\dfrac{\hbar^2k^2}{2m_{hh}}.~~~~~~~~~~~~~
\end{eqnarray}
As clear, the first two eigenvalues coincide with the band dispersion of ``conventional'' 3D topological insulators~\cite{TAI3Da,TAI3Db,TAI3Dc}, while the third branch represents an additional ``parabolic'' heavy-hole band. The mass parameter $M$ in Eq.~(\ref{eq:A8}) describes the band inversion: $M>0$ corresponds to a trivial semiconductor, while $M<0$ represents a topological semimetal. In the latter case, semi-metallic HgCdTe is often called ``semiconductor with negative band-gap'' assuming the energy difference between the $\Gamma_6$ and $\Gamma_8$ bands.

\begin{table*}[!]
\caption{\label{tab:1} Band parameters of bulk Hg$_{1-x}$Cd$_{x}$Te crystals at $T=2$~K calculated on the basis of material parameters provided in Ref.~\cite{th1}.}
\begin{ruledtabular}
\begin{tabular}{c|c|c|c|c|c|c|c}
x & ${m_{hh}/m_0}$ & $C$~(meV) & $M$~(meV) & $A$~(meV$\cdot$nm) & $B$~(meV$\cdot$nm$^2$) & $D$~(meV$\cdot$nm$^2$) & $a_0$ (nm) \\
\hline
0.155 & 0.341 & 0.00 & -12.17 & 691.02 & -103.30 & 66.26 & 0.645975 \\
$x_c\simeq$0.168 & 0.342 & 0.00 & 0.00 & 691.02 & -102.19 & 65.24 & 0.646003 \\
0.18 & 0.344 & 0.00 & 10.59 & 691.02 & -101.22 & 64.35 & 0.646027
\end{tabular}
\end{ruledtabular}
\end{table*}

In order to calculate DOS in bulk Hg$_{1-x}$Cd$_{x}$Te in the presence of disorder, we add the random impurity potential $V_{imp}(\textbf{r})I_{6{\times}6}$ and the ``gap fluctuation'' potential $U_M(\textbf{r})I_{M}$ to $H_{3D}(\mathbf{k})$. Further, we assume both potentials to be Gauss-distributed with zero mean values and define its spatial correlations as
\begin{eqnarray}
\label{eq:A11n}
\langle{V_{imp}(\textbf{r})}\rangle=\langle{U_M(\textbf{r})}\rangle=0,~~~~~\notag\\
\langle{V_{imp}(\textbf{r})V_{imp}(\textbf{r}')}\rangle=v^2g\left(\left|\textbf{r}-\textbf{r}'\right|\right),\notag\\
\langle{U_M(\textbf{r})U_M(\textbf{r}')}\rangle=u^2\xi\left(\left|\textbf{r}-\textbf{r}'\right|\right),\notag\\
\langle{V_{imp}(\textbf{r})U_M(\textbf{r}')}\rangle=0,~~~~~~
\end{eqnarray}
where $\langle...\rangle$ represents the average over all realizations of the random potentials; $g(r)$ are $\xi(r)$ are the normalized correlation functions~\cite{Int14,Int16}, while $v$ and $u$ represent the corresponding disorder strength. The last expression in Eq.~(\ref{eq:A11n}) means that the random potentials $V_{imp}(\textbf{r})$ and $U_M(\textbf{r})$ are supposed to be independent.

Let us consider Green's function defined by
\begin{equation}
\label{eq:A12}
\hat{G}(\mathbf{k},\varepsilon)=\langle\dfrac{1}{\varepsilon-\mathcal{H}}\rangle=
\left[\varepsilon-H_{3D}(\mathbf{k})-\hat{\Sigma}(\mathbf{k},\varepsilon)\right]^{-1},
\end{equation}
where $\varepsilon$ is the energy, $\hat{\Sigma}(\mathbf{k},\varepsilon)$ is the self-energy matrix, and $\mathcal{H}=H_{3D}(\mathbf{k})+V_{imp}(\textbf{r})I_{6{\times}6}+U_M(\textbf{r})I_{M}$.

Due to full rotational symmetry of $H_{3D}(\mathbf{k})$, the disorder-averaged Green's function in Eq.~(\ref{eq:A12}) can be presented in the form
\begin{equation}
\label{eq:A14}
\hat{G}(\mathbf{k},\varepsilon)=U_{z}(-\phi)U_{y}(-\theta)\hat{G}(k,\varepsilon)U_{y}(\theta)U_{z}(\phi),
\end{equation}
with
\begin{equation}
\label{eq:A15}
\hat{G}(k,\varepsilon)=\left[\varepsilon-\tilde{H}_{3D}(k)-\hat{\Sigma}(k,\varepsilon)\right]^{-1},
\end{equation}
which depends only on $k$. This shows that $\hat{G}(\mathbf{k},\varepsilon)$ in Eq.~(\ref{eq:A14}) depends on the angles via the terms of $U_{z}(\phi)$ and $U_{y}(\theta)$.

Without loss of generality, when calculating the self-energy $\hat{\Sigma}(\mathbf{k},\varepsilon)$, we further assume that the vector $\textbf{k}$ in its coordinate system is oriented along the $z$ axis. In this case, if one additionally takes into account that $\tilde{H}_{3D}(k)$ in Eq.~(\ref{eq:A14}) restricts the form of $\hat{\Sigma}(k,\varepsilon)$, one can write the self-energy $\hat{\Sigma}(k,\varepsilon)$ as
\begin{equation}
\label{eq:A17}
\hat{\Sigma}(k,\varepsilon)=\begin{pmatrix}
\Sigma_c & 0 & 0 & \Sigma_{p} & 0 & 0 \\
0 & \Sigma_c & 0 & 0 & \Sigma_{p} & 0 \\
0 & 0 & \Sigma_{hh} & 0 & 0 & 0 \\
\Sigma_{p} & 0 & 0 & \Sigma_{lh} & 0 & 0 \\
0 & \Sigma_{p} & 0 & 0 & \Sigma_{lh} & 0 \\
0 & 0 & 0 & 0 & 0 & \Sigma_{hh}
\end{pmatrix},
\end{equation}
where $\Sigma_c$, $\Sigma_{lh}$, $\Sigma_{hh}$ and $\Sigma_{p}$ are functions of $k$ and $\varepsilon$.

After some calculations involving Eqs.~(\ref{eq:A11n}--\ref{eq:A17}), the self-energy matrix $\hat{\Sigma}(k,\varepsilon)$ within the SCBA for correlated and \emph{mutually independent} disorder potentials~\cite{Int16,Int17,Int18} is written as
\begin{widetext}
\begin{multline}
\label{eq:A18}
\hat{\Sigma}(k,\varepsilon)=\int\limits_0^{K_c}\dfrac{{k^\prime}^{2}dk^\prime}{2\pi}
\int\limits_0^{\pi}\dfrac{\sin{\theta}d\theta}{2\pi}
W_{-}(k,k',\theta)G_{p}^\prime\cos{\theta}\begin{pmatrix}
0 & 0 & 0 & 1 & 0 & 0 \\
0 & 0 & 0 & 0 & 1 & 0 \\
0 & 0 & 0 & 0 & 0 & 0 \\
1 & 0 & 0 & 0 & 0 & 0 \\
0 & 1 & 0 & 0 & 0 & 0 \\
0 & 0 & 0 & 0 & 0 & 0
\end{pmatrix}+\\
+\int\limits_0^{K_c}\dfrac{{k^\prime}^{2}dk^\prime}{2\pi}
\int\limits_0^{\pi}\dfrac{\sin{\theta}d\theta}{2\pi}W_{+}(k,k',\theta)\left\{\begin{pmatrix}
G_{c}^\prime & 0 & 0 & 0 & 0 & 0 \\
0 & G_{c}^\prime & 0 & 0 & 0 & 0 \\
0 & 0 & G_{hh}^\prime & 0 & 0 & 0 \\
0 & 0 & 0 & G_{lh}^\prime & 0 & 0 \\
0 & 0 & 0 & 0 & G_{lh}^\prime & 0 \\
0 & 0 & 0 & 0 & 0 & G_{hh}^\prime
\end{pmatrix}
-\dfrac{3}{4}\left(G_{hh}^\prime-G_{lh}^\prime\right)\sin^{2}{\theta}\begin{pmatrix}
0 & 0 & 0 & 0 & 0 & 0 \\
0 & 0 & 0 & 0 & 0 & 0 \\
0 & 0 & 1 & 0 & 0 & 0 \\
0 & 0 & 0 & -1 & 0 & 0 \\
0 & 0 & 0 & 0 & -1 & 0 \\
0 & 0 & 0 & 0 & 0 & 1
\end{pmatrix}\right\},
\end{multline}
\end{widetext}
where $G_{n}^\prime\equiv G_{n}(k^\prime,\varepsilon)$ are the component of the Green's function $\hat{G}(k,\varepsilon)$ having the same form as $\hat{\Sigma}(k,\varepsilon)$ in Eq.~(\ref{eq:A17}), while $W_{\pm}(k,k',\theta)$ are defined as
\begin{multline}
\label{eq:A18n}
W_{\pm}(k,k',\theta)=v^2\tilde{g}\left(\sqrt{k^2+{k^\prime}^{2}-2{k}{k^\prime}\cos{\theta}}\right)
\pm\\
u^2\tilde{\xi}\left(\sqrt{k^2+{k^\prime}^{2}-2{k}{k^\prime}\cos{\theta}}\right),
\end{multline}
where $\tilde{g}(k)$ and $\tilde{\xi}(k)$ are the Fourier transform of the disorder correlation functions defined by Eq.~(\ref{eq:A11n}):
\begin{equation*}
g(r)=\int\dfrac{d^3\textbf{q}}{(2\pi)^3}e^{i\textbf{q}\cdot\textbf{r}}\tilde{g}(q),~~~~
\xi(r)=\int\dfrac{d^3\textbf{q}}{(2\pi)^3}e^{i\textbf{q}\cdot\textbf{r}}\tilde{\xi}(q).
\end{equation*}
In Eq.~(\ref{eq:A18}), we also introduce a cut-off wave-vector $K_c=\pi/a_{\mathrm{0}}$ (where $a_{\mathrm{0}}$ is the lattice constant of bulk semiconductor), corresponding to the size of the Brillouin zone~\cite{Int15}.

Once the Green's function is known, the spectral function $A(k,\varepsilon)$ and density-of-states $D(\varepsilon)$ are calculated as
\begin{eqnarray}
\label{eq:A19}
A(k,\varepsilon)=-\dfrac{1}{\pi}\textrm{Im}\left\{\textrm{Tr}\left(\hat{G}(k,\varepsilon+i0)\right)\right\},\notag\\
D(\varepsilon)=\int\limits_0^{K_c}\dfrac{k^2dk}{2{\pi}^2}A(k,\varepsilon).~~~~~~~~~~
\end{eqnarray}

\begin{figure*}
\includegraphics [width=2.05\columnwidth, keepaspectratio] {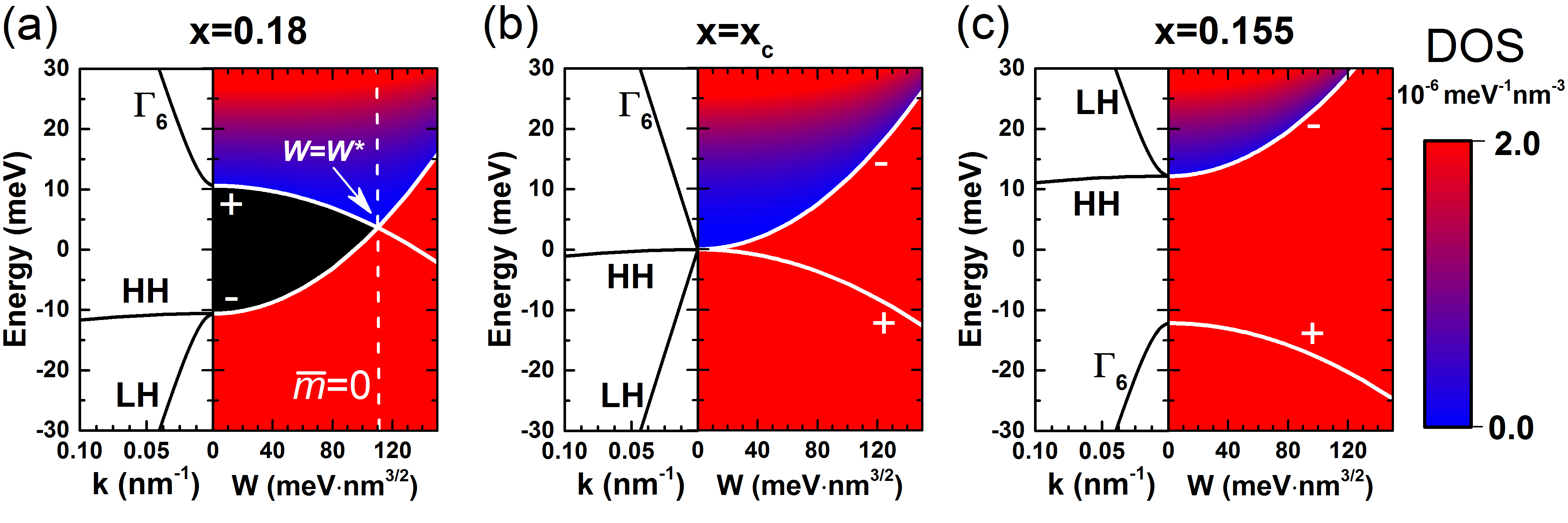} 
\caption{\label{Fig:1} Band structure and color map of the DOS as a function of the disorder strength~$W=\sqrt{v^2+u^2}$ calculated on the basis of three-band Kane Hamiltonian for bulk Hg$_{1-x}$Cd$_{x}$Te crystals at different Cd concentration: $x=0.18$ ($M>0$), $x=x_c$ ($M=0$ for $x_c\simeq0.168$ at 2~K~\cite{th1}) and $x=0.155$ ($M<0$). The band parameters are provided in Table~\ref{tab:1}. The white solid curves show the band edge positions determined by Eq.~(\ref{eq:A25}). The sign on the solid curves conforms to the corresponding sign in the formula. The white dotted curve found from $\overline{m}(\varepsilon,W)=0$ represents the topological phase transition.}
\end{figure*}
\begin{figure*}
\includegraphics [width=2.05\columnwidth, keepaspectratio] {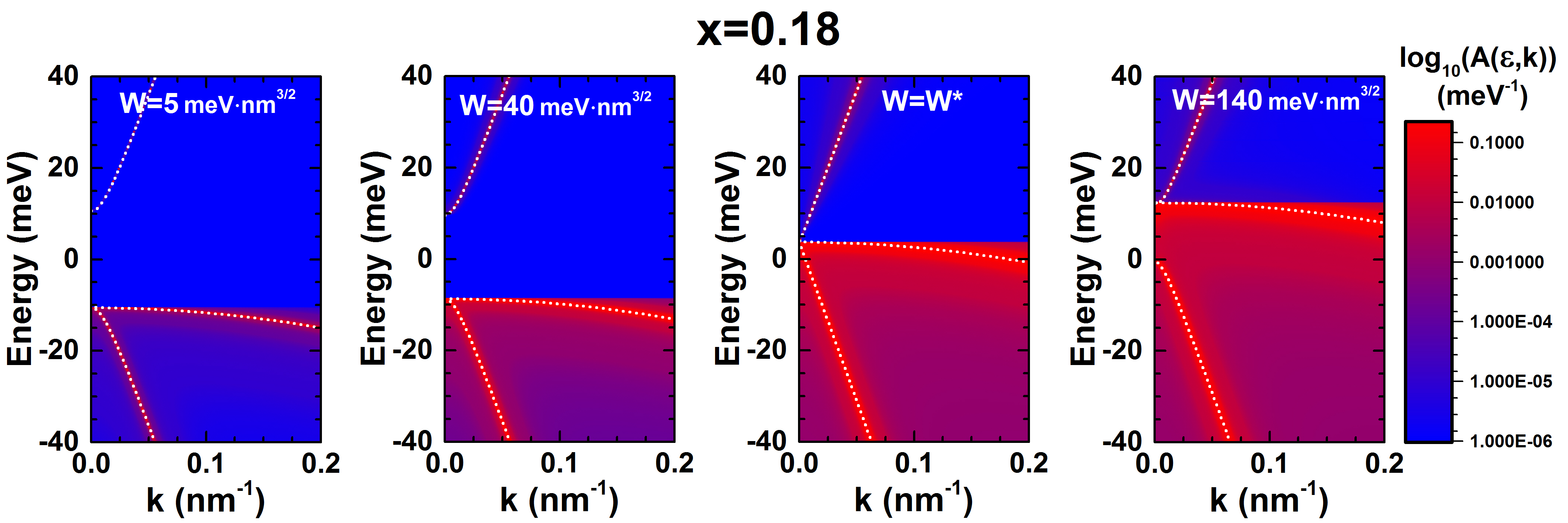} 
\caption{\label{Fig:2} Color map of the spectral function $A(k,\varepsilon)$ for bulk Hg$_{0.82}$Cd$_{0.18}$Te crystal at different values of the disorder strength~$W=\sqrt{v^2+u^2}$. The white dotted curves represent the quasiparticle energy dispersion defined by Eq.~(\ref{eq:A26}).}
\end{figure*}

To proceed further, we assume $\tilde{g}(q)=\tilde{\xi}(q)=1$, which corresponds to \emph{uncorrelated} disorder potentials $U_M(\textbf{r})$ and $V_{imp}(\textbf{r})$. The latter for instance can be considered as the electrostatic potential formed by the short-range impurities~\cite{Int15,Int19,Int20,Int21}. In this case, the self-energy matrix is independent of $k$ and has the diagonal form with $\Sigma_p(\varepsilon)=0$ and
\begin{eqnarray}
\label{eq:A20}
\Sigma_{hh}(\varepsilon)=\Sigma_{lh}(\varepsilon)=\dfrac{v^2+u^2}{2\pi^2}\int\limits_0^{K_c}\dfrac{G_{hh}(k,\varepsilon)+G_{lh}(k,\varepsilon)}{2}{k}^{2}d{k},\notag\\
\Sigma_c(\varepsilon)=\dfrac{v^2+u^2}{2\pi^2}\int\limits_0^{K_c}G_{c}(k,\varepsilon){k}^{2}d{k},~~~~~~~~~~~~~~~~
\end{eqnarray}
where $G_{c}(k,\varepsilon)$, $G_{hh}(k,\varepsilon)$ and $G_{lh}(k,\varepsilon)$ are written as
\begin{eqnarray}
\label{eq:A21}
G_{c}(k,\varepsilon)=\dfrac{(D-B)k^2+X_{lh}(\varepsilon)}{\Lambda(k,\varepsilon)},\notag\\
G_{lh}(k,\varepsilon)=\dfrac{(D+B)k^2+X_{c}(\varepsilon)}{\Lambda(k,\varepsilon)},\notag\\
G_{hh}(k,\varepsilon)=\dfrac{1}{\dfrac{\hbar^{2}k^{2}}{2m_{hh}}+X_{hh}(\varepsilon)}.~~~
\end{eqnarray}
Here, we introduce $X_{c}(\varepsilon)=\varepsilon-E_c-\Sigma_c(\varepsilon)$, $X_{lh}(\varepsilon)=\varepsilon-E_v-\Sigma_{lh}(\varepsilon)$, $X_{hh}(\varepsilon)=\varepsilon-E_v-\Sigma_{hh}(\varepsilon)$ and
\begin{multline}
\label{eq:A22}
\Lambda(k,\varepsilon)=\left[D^2-B^2\right]k^4-\\
-\left[A^2-(D-B)X_c-(D+B)X_{lh}\right]k^2+X_cX_{lh}.
\end{multline}
Note that the integrals in Eq.~(\ref{eq:A20}) are calculated analytically (see Appendix~\ref{sec:Int}), transforming Eqs~(\ref{eq:A20}-\ref{eq:A22}) into the set of algebraic self-consistent equations numerically solved by simple iterations.

After the Green's function $\hat{G}(k,\varepsilon)$ and self-energy matrix $\hat{\Sigma}(k,\varepsilon)$ are known, the spectral function $A(k,\varepsilon)$ and density-of-states $D(\varepsilon)$ are calculated as
\begin{eqnarray}
\label{eq:A23}
A(k,\varepsilon)=-\dfrac{2}{\pi}\textrm{Im}\left\{G_{c}\left(k,\varepsilon\right)
+G_{lh}\left(k,\varepsilon\right)+G_{hh}\left(k,\varepsilon\right)\right\},\notag\\
D(\varepsilon)=-\dfrac{2}{\pi}
\dfrac{\textrm{Im}\left\{\Sigma_{c}(\varepsilon)
+\Sigma_{lh}(\varepsilon)+\Sigma_{hh}(\varepsilon)\right\}}{v^2+u^2}
.~~~~~~~~~
\end{eqnarray}
Note that these expressions are valid only for the case of \emph{uncorrelated} disorder potentials $U_M(\textbf{r})$ and $V_{imp}(\textbf{r})$.

\section{\label{Sec:RnD} Results and Discussion}
Figure~\ref{Fig:1} shows the evolution of DOS with the disorder strength $W$ introduced as $W=\sqrt{v^2+u^2}$ in accordance with Eq.~(\ref{eq:A20}) for Hg$_{1-x}$Cd$_{x}$Te crystals with different Cd concentration $x$. If $M>0$, the band-gap in the DOS decreases with $W$ until it vanishes above a critical value $W^{*}$. Conversely, the DOS at $M<0$ remains gapless for all values of $W$. Let us now demonstrate that such behavior of the DOS in CdHgTe crystals is attributed to the disorder-induced topological phase transition.

Indeed, according to Groth~\emph{et~al.}~\cite{Int10}, the presence of disorder leads to the renormalization of both topological mass $\overline{m}(\varepsilon,W)$ and chemical potential $\overline{\mu}(\varepsilon,W)$:
\begin{eqnarray}
\label{eq:A24}
\overline{m}(\varepsilon,W)=M+\dfrac{\overline{\Sigma}_{c}(\varepsilon)-\overline{\Sigma}_{lh}(\varepsilon)}{2},\notag\\
\overline{\mu}(\varepsilon,W)=C-\dfrac{\overline{\Sigma}_{c}(\varepsilon)+\overline{\Sigma}_{lh}(\varepsilon)}{2},~
\end{eqnarray}
where the overbar stresses the values found on the set of real numbers. Since the finite DOS is associated with a finite imaginary part of the self-energy matrix $\hat{\Sigma}(\varepsilon)$, the band-gap region is characterized by the solution of Eq.~(\ref{eq:A20}) with purely real quantities $\overline{\Sigma}_{c}(\varepsilon)$, $\overline{\Sigma}_{lh}(\varepsilon)$ and $\overline{\Sigma}_{hh}(\varepsilon)$. Note that $\overline{\Sigma}_{hh}(\varepsilon)=\overline{\Sigma}_{lh}(\varepsilon)$ for the Kane fermions. A topological transition occurs when the renormalized mass parameter changes the sign, i.e. $\overline{m}(\varepsilon,W)=0$~\cite{Int10}. The band edge positions on the ($\varepsilon$,$W$)-diagram are found from the condition:
\begin{equation}
\label{eq:A25}
\overline{\mu}(\varepsilon,W)=C\pm\overline{m}(\varepsilon,W),
\end{equation}
where ``$+$'' and ``-'' correspond to the edge of $\Gamma_6$ and $\Gamma_8$ bands, respectively.

As seen from Fig.~\ref{Fig:1}(a), two curves corresponding to the band edges cross at the transition point $W=W^{*}$, where $\overline{m}(\varepsilon,W)=0$. Interestingly, these curves not only describe the boundaries of the region with vanishing DOS, but also allow one to trace the band edge positions in the absence of the band-gap at $\overline{m}(\varepsilon,W)<0$. The latter is also seen in Fig.~\ref{Fig:1}(b,c). The description of disorder-induced phase transition in terms of the renormalized topological mass and chemical potential implies the implicit replacement of Kane fermions by \emph{new quasiparticles} with energy dispersion determined by
\begin{equation}
\label{eq:A26}
\mathrm{det}\left|\tilde{H}_{3D}(k)+\hat{\overline{\Sigma}}(\varepsilon)-{\varepsilon}I_{6{\times}6}\right|=0,
\end{equation}
where the self-energy matrix is found on the set of real numbers.

\begin{figure*}
\includegraphics [width=2.05\columnwidth, keepaspectratio] {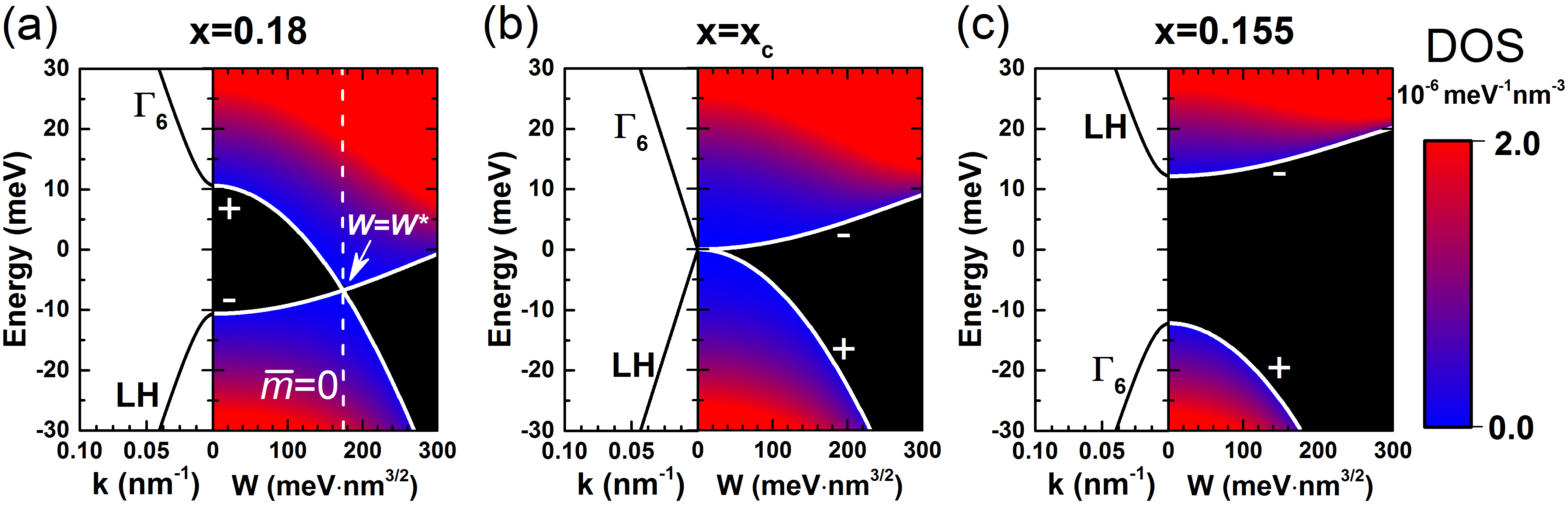} 
\caption{\label{Fig:3} Band structure and color map of the DOS as a function of the disorder strength~$W=\sqrt{v^2+u^2}$ calculated within the two-band 3D BHZ model (see Appendix~\ref{sec:BHZ}) with the same band parameters as used for Fig.~\ref{Fig:1}. The white solid curves show the band edge positions determined by Eq.~(\ref{eq:A25}). The sign on the solid curves conforms to the corresponding sign in the formula. The dotted curve found from $\overline{m}(\varepsilon,W)=0$ represents the topological phase transition.}
\end{figure*}

In order to shed further light on this issue, Fig.~\ref{Fig:2} provides color maps of the spectral function $A(k,\varepsilon)$ for bulk Hg$_{0.82}$Cd$_{0.18}$Te crystal at different values of the disorder strength. As clear, at $W=0$ corresponding to the ``clean'' crystal, $\hat{\Sigma}(\varepsilon){\equiv}0$ and the spectral function is represented by sum of $\delta$-functions centered at the energies $E_{c}$, $E_{lh}$ and $E_{hh}$ given by Eq.~(\ref{eq:A10}). Figure~\ref{Fig:2} evidences that although the spectral function broadens in the presence of disorder, its maximum values still coincides with the energy dispersion of quasiparticles defined by Eq.~(\ref{eq:A26}) and shown by the white dotted curves. As also seen from the evolution of $A(k,\varepsilon)$, the energy dispersion of quasiparticles mimics the energy dispersion of Kane fermions with the mass parameter renormalized by disorder. This supports the quasiparticle picture of the disorder-induced phase transition. Thus, the mechanism previously discovered by Groth~\emph{et~al.}~\cite{Int10} for Dirac systems with quadratic momentum terms also takes place in narrow-gap HgCdTe systems described by the three-band Kane Hamiltonian.

Nevertheless, it is important to underline the non-Lorentzian shape of $A(k,\varepsilon)$ in disordered CdHgTe crystals. This signalizing that no perfectly coherent quasiparticles can be defined, and incoherent processes associated with the imaginary parts of the self-energy are relevant in the presence of disorder. The strongly asymmetrical shape of the spectral function is particularly seen in the evolution of heavy-hole branch.

Finally, let us discuss the contribution of heavy-hole band into disorder-induced phase transition. As seen from Eqs.~(\ref{eq:A20})--(\ref{eq:A22}), all three bands are self-consistently involved in the DOS calculations, which means that the heavy-hole band cannot be considered separately. However, we can compare the DOS evolution shown in Fig.~\ref{Fig:1} with the calculations on the basis of two-band 3D BHZ Hamiltonian, whose eigenvalues formally coincide with the dispersion of the $\Gamma_6$ and light-hole bands. Details of these calculations are provided in Appendix~\ref{sec:BHZ}.

Figure~\ref{Fig:3} shows the evolution of DOS with the disorder strength $W$ calculated within the two-band 3D BHZ model (see Appendix~\ref{sec:BHZ}) with the same band parameters as used for Fig.~\ref{Fig:1}. If $M>0$, the band-gap decreases with $W$ and vanishes at a critical value $W^{*}$, and then it is reopened again at $W>W^{*}$. Such behavior reminds ``typical'' phase transition between trivial and topological insulator states known for HgTe QWs~\cite{Int10,Int15}. In the absence of heavy-hole band, the disorder induces a phase transition between trivial and topological insulator states. The presence of heavy-hole band in HgCdTe crystals leads to the topological phase transition into the semimetal state, which occurs at much lower disorder strength $W^{*}$ than the transition in conventional 3D topological insulators (cf.~Fig.~\ref{Fig:1}).

\begin{figure}
\includegraphics [width=1.00\columnwidth, keepaspectratio] {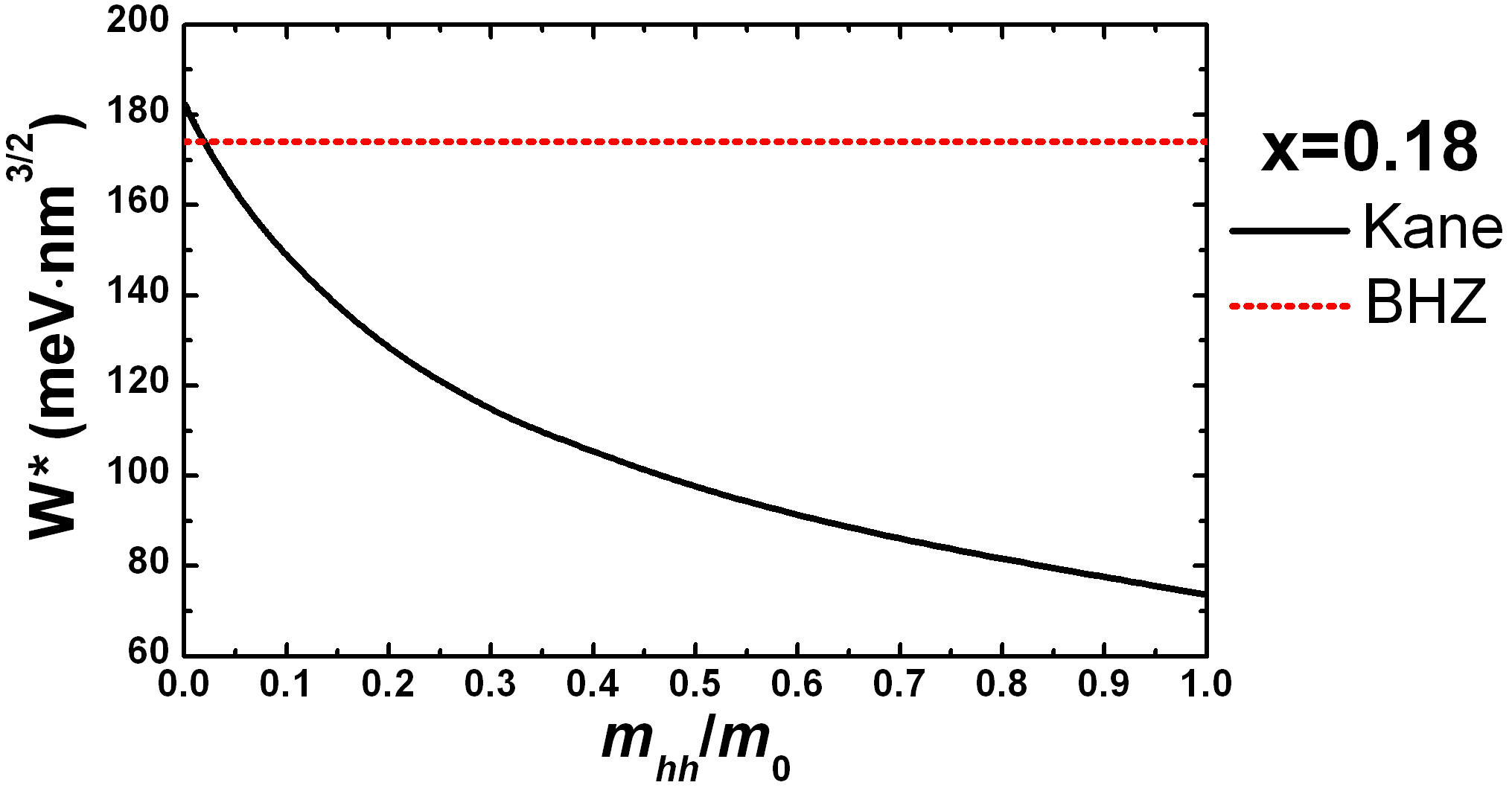} 
\caption{\label{Fig:4} Critical disorder strength $W^{*}$ as a function of $m_{hh}$ for bulk Hg$_{0.82}$Cd$_{0.18}$Te. The red dotted curve represents the value calculated within the two-band 3D BHZ model shown in Fig.~\ref{Fig:3}(a).}
\end{figure}

The difference between the values of $W^{*}$ calculated in the two models is not surprising. Although the unitary transformation~(\ref{eq:A6}) indeed allows to represent $H_{3D}(\mathbf{k})$ in the form in which the heavy-holes are completely decoupled from the electron and light-hole bands, the elements of self-energy matrix $\hat{\Sigma}(k,\varepsilon)$ in Eq.~(\ref{eq:A20}) still include the mixed contribution from all three bands. Indeed, in order to calculate $\Sigma_c$ via $G_c$, one needs to know $\Sigma_{lh}$, which in turn depends on both $G_{lh}$ and $G_{hh}$. The inability to separate the light and heavy-holes, when calculating the Green function, is due to the fact that these two bands are described within the same representation of $J=3/2$. As for the fact that the value $W^{*}$ lower than the calculated one by using the two-band 3D BHZ model (see Fig.~\ref{Fig:3}(a)), this is due to the actual values of heavy-hole mass $m_{hh}$ in HgCdTe crystals (see Table~\ref{tab:1}). Figure~\ref{Fig:4}, providing $W^{*}$ as a function of $m_{hh}$, shows that the critical disorder strength at small $m_{hh}$ values exceeds $W^{*}$ calculated by using the two-band 3D BHZ model, while increasing of $m_{hh}$ leads to decreasing of $W^{*}$.

\section{\label{Sec:Sum} Summary}
We have investigated disorder-induced topological phase transition in bulk CdHgTe crystals, in which the band structure is represented by specific Kane fermions differ from pseudo-spin-1/2 fermionic excitations. By using the SCBA and three-band Kane Hamiltonian in the \emph{continuous} representation, we directly calculate the DOS as a function of the strength of \emph{uncorrelated} disorder in HgCdTe crystals. By following the DOS evolution, we clearly demonstrate the topological phase transition, which can be also understood within the quasiparticle picture in terms of the disorder-renormalized mass of Kane fermions. Our conclusions also hold for other narrow-gap zinc-blende semiconductors such as InAs, InSb and their ternary alloys InAsSb.

In this work, as a source of the disorder, we have considered the electrostatic potential $V_{imp}(\textbf{r})$ of randomly distributed \emph{short-range} impurities and \emph{short-range} fluctuations $U_M(\textbf{r})$ in the Cd composition for which $\langle{V_{imp}(\textbf{r})U_M(\textbf{r}')}\rangle=0$. Nevertheless, our results can be also generalized for other disorder models inherent in realistic crystals. This may be significant for the investigations aimed at improving the performance of InAsSb- and HgCdTe-based mid-infrared detectors~\cite{Int22}.

\begin{acknowledgments}
This work was supported by the Terahertz Occitanie Platform, by the Centre National de la Recherche Scientifique through IRP ``TeraMIR'' and the French Agence Nationale pour la Recherche (``Colector'' project).
\end{acknowledgments}

\appendix
\section{\label{sec:Int} Calculation of integrals}
In order to calculate the integrals in Eq.~(\ref{eq:A20}), we have found the roots
$x_1$, $x_2$, of the polynomial $\Lambda(k,\varepsilon)$ in Eq.~(\ref{eq:A22})
defining the following expansion: $\Lambda(x,\varepsilon)=\left[D^2-B^2\right]\left(x^2-x_1^2\right)\left(x^2-x_2^2\right)$.
Once the roots are known, the self-energies $\Sigma_{c}(\varepsilon)$, $\Sigma_{lh}(\varepsilon)$ and $\Sigma_{hh}(\varepsilon)$  in Eq.~(\ref{eq:A20}) are reduced to the calculation of the integrals
\begin{multline}
\label{eq:App1}
\int\dfrac{x^2\left(ax^2+b\right)}{c\left(x^2-x_1^2\right)\left(x^2-x_2^2\right)}dx=\\
=\dfrac{a}{c}x+\dfrac{x_1\left(ax_1^2+b\right)}{2c\left(x_1^2-x_2^2\right)}\left\{\ln\left(x-x_1\right)-\ln\left(x+x_1\right)\right\}-\\
-\dfrac{x_2\left(ax_2^2+b\right)}{2c\left(x_1^2-x_2^2\right)}\left\{\ln\left(x-x_2\right)-\ln\left(x+x_2\right)\right\}.
\end{multline}
and
\begin{equation}
\label{eq:App2}
\int\dfrac{ax^2}{c\left(x^2-x_3^2\right)}dx
=\dfrac{a}{c}x+\dfrac{ax_3}{2c}\left\{\ln\left(x-x_3\right)-\ln\left(x+x_3\right)\right\},
\end{equation}
where $x_3^2=-2m_{hh}X_{hh}(\varepsilon)/\hbar^2$.

\section{\label{sec:BHZ} 3D isotropical BHZ model and SCBA}
For better understanding of the specifics of HgCdTe crystals in comparison with conventional 3D topological insulators~\cite{TAI3Da,TAI3Db,TAI3Dc}, let us consider the disorder-induced phase transition within \emph{continuous} 3D BHZ model~\cite{th4,th5,th6}.
Formally, this two-band Hamiltonian describes a model system consisting of conduction and light-hole bands only, while the heavy-hole band is formally absent. In the basis set of $\left\{|\Gamma_6,+1/2\rangle,|\Gamma_6,-1/2\rangle,|\Gamma_8,+1/2\rangle,|\Gamma_8,-1/2\rangle\right\}$, the 3D BHZ Hamiltonian is written as~\cite{th4,th5,th6}:
\begin{multline}
\label{eq:ApB1}
H_{\mathrm{BHZ}}(\mathbf{k})=\left[C-D(k_x^2+k_y^2+k_z^2)\right]I_{4{\times}4}+\\
+\begin{pmatrix}
\mathcal{M}(\mathbf{k}) & 0 & Ak_{z} & Ak_{-} \\
0 & \mathcal{M}(\mathbf{k}) & Ak_{+} & -Ak_{z} \\
Ak_{z} & Ak_{-} & -\mathcal{M}(\mathbf{k}) & 0 \\
Ak_{+} & -Ak_{z} & 0 & -\mathcal{M}(\mathbf{k})
\end{pmatrix},
\end{multline}
where $\mathcal{M}(\mathbf{k})=M-Bk^2$ with $k^2=k_x^2+k_y^2+k_z^2$. Here, the structural parameters $C$, $M$, $B$, $D$ and $A$ are determined in the same way as those for Eq.~(\ref{eq:A8}). As clear, $H_{\mathrm{BHZ}}(\mathbf{k})$ has two doubly-degenerate eigenvalues
\begin{equation}
\label{eq:ApB2}
E_{c,lh}=C-Dk^2
\pm\sqrt{\left(M-Bk^2\right)^2+A^2k^2},
\end{equation}
which formally coincide with the band dispersion within the three-band Kane Hamiltonian (see Eq.~(\ref{eq:A10})).

The full rotational symmetry of $H_{\mathrm{BHZ}}(\mathbf{k})$ allows for the representation of its wave-function $\Psi_{\mathrm{BHZ}}(\mathbf{k})$ in the form:
\begin{equation}
\label{eq:ApB3}
\Psi_{\mathrm{BHZ}}(\mathbf{k})=T_{z}(\phi)T_{y}(\theta)\Psi_{\mathrm{BHZ}}(k),
\end{equation}
with
\begin{eqnarray}
\label{eq:ApB4}
T_{z}(\phi)=\begin{pmatrix}
\exp\left(-i\sigma_z\phi/2\right) & 0 \\0 & \exp\left(-i\sigma_z\phi/2\right)
\end{pmatrix},\notag\\
T_{y}(\theta)=\begin{pmatrix}
\exp(-i\sigma_y\theta/2) & 0 \\0 & \exp(-i\sigma_y\theta/2)
\end{pmatrix},~
\end{eqnarray}
where $k_x=k\sin{\theta}\cos{\phi}$, $k_y=k\sin{\theta}\sin{\phi}$, $k_z=k\cos{\theta}$.

The unitary transformations defined by $T_{z}(\phi)$ and $T_{y}(\theta)$ allows one to introduce a new Hamiltonian $\tilde{H}_{\mathrm{BHZ}}(k)=T_{y}(-\theta)T_{z}(-\phi)H_{\mathrm{BHZ}}(\mathbf{k})T_{z}(\phi)T_{y}(\theta)$, which has a form:
\begin{multline}
\label{eq:ApB5}
\tilde{H}_{\mathrm{BHZ}}(k)=\left[C-Dk^2\right]I_{4{\times}4}+\\
+\begin{pmatrix}
\left[M-Bk^2\right]I_{2{\times}2} & Ak\sigma_{z} \\
Ak\sigma_{z} & -\left[M-Bk^2\right]I_{2{\times}2}
\end{pmatrix}.
\end{multline}

Then, with the unitary transformation of Eq.~(\ref{eq:ApB3}), the disorder-averaged Green's function $\hat{G}_{\mathrm{BHZ}}(\mathbf{k},\varepsilon)$ can be presented in the form
\begin{equation}
\label{eq:ApB6}
\hat{G}_{\mathrm{BHZ}}(\mathbf{k},\varepsilon)=T_{z}(-\phi)T_{y}(-\theta)\hat{G}_{\mathrm{BHZ}}(k,\varepsilon)T_{y}(\theta)T_{z}(\phi),
\end{equation}
with
\begin{equation}
\label{eq:ApB7}
\hat{G}_{\mathrm{BHZ}}(k,\varepsilon)=\left[\varepsilon-\tilde{H}_{\mathrm{BHZ}}(k)-\hat{\Sigma}_{\mathrm{BHZ}}(k,\varepsilon)\right]^{-1},
\end{equation}
where the self-energy $\hat{\Sigma}_{\mathrm{BHZ}}(k,\varepsilon)$ has a form of $\tilde{H}_{\mathrm{BHZ}}(k)$:
\begin{equation}
\label{eq:ApB8}
\hat{\Sigma}_{\mathrm{BHZ}}(k,\varepsilon)=\begin{pmatrix}
{\Sigma}_{c}^{\mathrm{(BHZ)}}I_{2{\times}2} & {\Sigma}_{p}^{\mathrm{(BHZ)}}\sigma_{z} \\
{\Sigma}_{p}^{\mathrm{(BHZ)}}\sigma_{z} & {\Sigma}_{lh}^{\mathrm{(BHZ)}}I_{2{\times}2}
\end{pmatrix}.
\end{equation}
Then, after some calculations similar to the case of Kane fermions, the self-energy matrix $\hat{\Sigma}_{\mathrm{BHZ}}(k,\varepsilon)$ within the SCBA for the Gauss-distributed disorder potentials in Eq.~(\ref{eq:A11n}) is written as
\begin{multline}
\label{eq:ApB9}
\hat{\Sigma}_{\mathrm{BHZ}}(k,\varepsilon)=\int\limits_0^{K_c}\dfrac{{k^\prime}^{2}dk^\prime}{2\pi}
\int\limits_0^{\pi}\dfrac{\sin{\theta}d\theta}{2\pi}\cdot\\
\cdot\begin{pmatrix}
W_{+}{G_{c}^{\mathrm{(BHZ)}}}^{\prime}I_{2{\times}2} & W_{-}{G_{p}^{\mathrm{(BHZ)}}}^{\prime}\cos{\theta}\sigma_{z} \\
W_{-}{G_{p}^{\mathrm{(BHZ)}}}^{\prime}\cos{\theta}\sigma_{z} & W_{+}{G_{lh}^{\mathrm{(BHZ)}}}^{\prime}I_{2{\times}2}
\end{pmatrix},
\end{multline}
where $W_{\pm}\equiv W_{\pm}(k,k',\theta)$ are defined by Eq.~(\ref{eq:A18n}), while ${G_{c}^{\mathrm{(BHZ)}}}^{\prime}\equiv G_{c}^{\mathrm{(BHZ)}}(k^\prime,\varepsilon)$ are the component of the Green's function $\hat{G}_{\mathrm{BHZ}}(\mathbf{k},\varepsilon)$ having the same form as $\hat{\Sigma}_{\mathrm{BHZ}}(k,\varepsilon)$ in Eq.~(\ref{eq:ApB8}).

In the case of \emph{uncorrelated} disorder, when $W_{\pm}(k,k',\theta)=v^2{\pm}u^2$, the self-energy matrix is independent of $k$ and is written in the form:
\begin{eqnarray}
\label{eq:ApB10}
\Sigma_{c}^{\mathrm{(BHZ)}}(\varepsilon)=\dfrac{v^2+u^2}{2\pi^2}\int\limits_0^{K_c}\dfrac{(D-B)k^2+X_{lh}(\varepsilon)}{\Lambda(k,\varepsilon)}{k}^{2}d{k},\notag\\
\Sigma_{lh}^{\mathrm{(BHZ)}}(\varepsilon)=\dfrac{v^2+u^2}{2\pi^2}\int\limits_0^{K_c}\dfrac{(D+B)k^2+X_{c}(\varepsilon)}{\Lambda(k,\varepsilon)}{k}^{2}d{k},\notag\\
\Sigma_{p}^{\mathrm{(BHZ)}}(\varepsilon)=0,~~~~~~~~~~~~~~~~~~~~~~~~
\end{eqnarray}
where $X_{c}(\varepsilon)=\varepsilon-E_c-\Sigma_c(\varepsilon)$, $X_{lh}(\varepsilon)=\varepsilon-E_v-\Sigma_{lh}(\varepsilon)$, $X_{hh}(\varepsilon)=\varepsilon-E_v-\Sigma_{hh}(\varepsilon)$ and
\begin{multline*}
\Lambda(k,\varepsilon)=\left[D^2-B^2\right]k^4-\\
-\left[A^2-(D-B)X_c-(D+B)X_{lh}\right]k^2+X_cX_{lh}.
\end{multline*}

After the self-energy matrix $\hat{\Sigma}(k,\varepsilon)$ is known, the density-of-states $D(\varepsilon)$ is calculated as
\begin{equation}
\label{eq:ApB11}
D^{\mathrm{(BHZ)}}(\varepsilon)=-\dfrac{2}{\pi}
\dfrac{\textrm{Im}\left\{\Sigma_{c}(\varepsilon)+\Sigma_{lh}(\varepsilon)\right\}}{v^2+u^2}.
\end{equation}
One can see that Eqs~(\ref{eq:ApB10}) and (\ref{eq:ApB11}) differ from those for the Kane fermions (see Sec.~\ref{Sec:Theory}).


\begin{thebibliography}{41}%
\makeatletter
\providecommand \@ifxundefined [1]{%
 \@ifx{#1\undefined}
}%
\providecommand \@ifnum [1]{%
 \ifnum #1\expandafter \@firstoftwo
 \else \expandafter \@secondoftwo
 \fi
}%
\providecommand \@ifx [1]{%
 \ifx #1\expandafter \@firstoftwo
 \else \expandafter \@secondoftwo
 \fi
}%
\providecommand \natexlab [1]{#1}%
\providecommand \enquote  [1]{``#1''}%
\providecommand \bibnamefont  [1]{#1}%
\providecommand \bibfnamefont [1]{#1}%
\providecommand \citenamefont [1]{#1}%
\providecommand \href@noop [0]{\@secondoftwo}%
\providecommand \href [0]{\begingroup \@sanitize@url \@href}%
\providecommand \@href[1]{\@@startlink{#1}\@@href}%
\providecommand \@@href[1]{\endgroup#1\@@endlink}%
\providecommand \@sanitize@url [0]{\catcode `\\12\catcode `\$12\catcode
  `\&12\catcode `\#12\catcode `\^12\catcode `\_12\catcode `\%12\relax}%
\providecommand \@@startlink[1]{}%
\providecommand \@@endlink[0]{}%
\providecommand \url  [0]{\begingroup\@sanitize@url \@url }%
\providecommand \@url [1]{\endgroup\@href {#1}{\urlprefix }}%
\providecommand \urlprefix  [0]{URL }%
\providecommand \Eprint [0]{\href }%
\providecommand \doibase [0]{http://dx.doi.org/}%
\providecommand \selectlanguage [0]{\@gobble}%
\providecommand \bibinfo  [0]{\@secondoftwo}%
\providecommand \bibfield  [0]{\@secondoftwo}%
\providecommand \translation [1]{[#1]}%
\providecommand \BibitemOpen [0]{}%
\providecommand \bibitemStop [0]{}%
\providecommand \bibitemNoStop [0]{.\EOS\space}%
\providecommand \EOS [0]{\spacefactor3000\relax}%
\providecommand \BibitemShut  [1]{\csname bibitem#1\endcsname}%
\let\auto@bib@innerbib\@empty
\bibitem [{\citenamefont {Orlita}\ \emph {et~al.}(2014)\citenamefont {Orlita},
  \citenamefont {Basko}, \citenamefont {Zholudev}, \citenamefont {Teppe},
  \citenamefont {Knap}, \citenamefont {Gavrilenko}, \citenamefont {Mikhailov},
  \citenamefont {Dvoretskii}, \citenamefont {Neugebauer}, \citenamefont
  {Faugeras}, \citenamefont {Barra}, \citenamefont {Martinez},\ and\
  \citenamefont {Potemski}}]{Int1}%
  \BibitemOpen
  \bibfield  {author} {\bibinfo {author} {\bibfnamefont {M.}~\bibnamefont
  {Orlita}}, \bibinfo {author} {\bibfnamefont {D.~M.}\ \bibnamefont {Basko}},
  \bibinfo {author} {\bibfnamefont {M.~S.}\ \bibnamefont {Zholudev}}, \bibinfo
  {author} {\bibfnamefont {F.}~\bibnamefont {Teppe}}, \bibinfo {author}
  {\bibfnamefont {W.}~\bibnamefont {Knap}}, \bibinfo {author} {\bibfnamefont
  {V.~I.}\ \bibnamefont {Gavrilenko}}, \bibinfo {author} {\bibfnamefont
  {N.~N.}\ \bibnamefont {Mikhailov}}, \bibinfo {author} {\bibfnamefont {S.~A.}\
  \bibnamefont {Dvoretskii}}, \bibinfo {author} {\bibfnamefont
  {P.}~\bibnamefont {Neugebauer}}, \bibinfo {author} {\bibfnamefont
  {C.}~\bibnamefont {Faugeras}}, \bibinfo {author} {\bibfnamefont {A.-L.}\
  \bibnamefont {Barra}}, \bibinfo {author} {\bibfnamefont {G.}~\bibnamefont
  {Martinez}}, \ and\ \bibinfo {author} {\bibfnamefont {M.}~\bibnamefont
  {Potemski}},\ }\href {\doibase 10.1038/nphys2857} {\bibfield  {journal}
  {\bibinfo  {journal} {Nature Phys.}\ }\textbf {\bibinfo {volume} {10}},\
  \bibinfo {pages} {233} (\bibinfo {year} {2014})}\BibitemShut {NoStop}%
\bibitem [{\citenamefont {Teppe}\ \emph {et~al.}(2016)\citenamefont {Teppe},
  \citenamefont {Marcinkiewicz}, \citenamefont {Krishtopenko}, \citenamefont
  {Ruffenach}, \citenamefont {Consejo}, \citenamefont {Kadykov}, \citenamefont
  {Desrat}, \citenamefont {But}, \citenamefont {Knap}, \citenamefont {Ludwig},
  \citenamefont {Moon}, \citenamefont {Smirnov}, \citenamefont {Orlita},
  \citenamefont {Jiang}, \citenamefont {Morozov}, \citenamefont {Gavrilenko},
  \citenamefont {Mikhailov},\ and\ \citenamefont {Dvoretskii}}]{Int2}%
  \BibitemOpen
  \bibfield  {author} {\bibinfo {author} {\bibfnamefont {F.}~\bibnamefont
  {Teppe}}, \bibinfo {author} {\bibfnamefont {M.}~\bibnamefont
  {Marcinkiewicz}}, \bibinfo {author} {\bibfnamefont {S.~S.}\ \bibnamefont
  {Krishtopenko}}, \bibinfo {author} {\bibfnamefont {S.}~\bibnamefont
  {Ruffenach}}, \bibinfo {author} {\bibfnamefont {C.}~\bibnamefont {Consejo}},
  \bibinfo {author} {\bibfnamefont {A.~M.}\ \bibnamefont {Kadykov}}, \bibinfo
  {author} {\bibfnamefont {W.}~\bibnamefont {Desrat}}, \bibinfo {author}
  {\bibfnamefont {D.}~\bibnamefont {But}}, \bibinfo {author} {\bibfnamefont
  {W.}~\bibnamefont {Knap}}, \bibinfo {author} {\bibfnamefont {J.}~\bibnamefont
  {Ludwig}}, \bibinfo {author} {\bibfnamefont {S.}~\bibnamefont {Moon}},
  \bibinfo {author} {\bibfnamefont {D.}~\bibnamefont {Smirnov}}, \bibinfo
  {author} {\bibfnamefont {M.}~\bibnamefont {Orlita}}, \bibinfo {author}
  {\bibfnamefont {Z.}~\bibnamefont {Jiang}}, \bibinfo {author} {\bibfnamefont
  {S.~V.}\ \bibnamefont {Morozov}}, \bibinfo {author} {\bibfnamefont
  {V.}~\bibnamefont {Gavrilenko}}, \bibinfo {author} {\bibfnamefont {N.~N.}\
  \bibnamefont {Mikhailov}}, \ and\ \bibinfo {author} {\bibfnamefont {S.~A.}\
  \bibnamefont {Dvoretskii}},\ }\href {\doibase 10.1038/ncomms12576} {\bibfield
   {journal} {\bibinfo  {journal} {Nat. Commun.}\ }\textbf {\bibinfo {volume}
  {7}},\ \bibinfo {pages} {12576} (\bibinfo {year} {2016})}\BibitemShut
  {NoStop}%
\bibitem [{\citenamefont {But}\ \emph {et~al.}(2019)\citenamefont {But},
  \citenamefont {Mittendorff}, \citenamefont {Consejo}, \citenamefont {Teppe},
  \citenamefont {Mikhailov}, \citenamefont {Dvoretskii}, \citenamefont
  {Faugeras}, \citenamefont {Winnerl}, \citenamefont {Helm}, \citenamefont
  {Knap}, \citenamefont {Potemski},\ and\ \citenamefont {Orlita}}]{Int2b}%
  \BibitemOpen
  \bibfield  {author} {\bibinfo {author} {\bibfnamefont {D.}~\bibnamefont
  {But}}, \bibinfo {author} {\bibfnamefont {M.}~\bibnamefont {Mittendorff}},
  \bibinfo {author} {\bibfnamefont {C.}~\bibnamefont {Consejo}}, \bibinfo
  {author} {\bibfnamefont {F.}~\bibnamefont {Teppe}}, \bibinfo {author}
  {\bibfnamefont {N.}~\bibnamefont {Mikhailov}}, \bibinfo {author}
  {\bibfnamefont {S.~A.}\ \bibnamefont {Dvoretskii}}, \bibinfo {author}
  {\bibfnamefont {C.}~\bibnamefont {Faugeras}}, \bibinfo {author}
  {\bibfnamefont {S.}~\bibnamefont {Winnerl}}, \bibinfo {author} {\bibfnamefont
  {M.}~\bibnamefont {Helm}}, \bibinfo {author} {\bibfnamefont {W.}~\bibnamefont
  {Knap}}, \bibinfo {author} {\bibfnamefont {M.}~\bibnamefont {Potemski}}, \
  and\ \bibinfo {author} {\bibfnamefont {M.}~\bibnamefont {Orlita}},\ }\href
  {\doibase 10.1038/s41566-019-0496-1} {\bibfield  {journal} {\bibinfo
  {journal} {Nat. Photonics}\ }\textbf {\bibinfo {volume} {13}},\ \bibinfo
  {pages} {783} (\bibinfo {year} {2019})}\BibitemShut {NoStop}%
\bibitem [{\citenamefont {Krishtopenko}\ and\ \citenamefont
  {Teppe}(2022)}]{Int3}%
  \BibitemOpen
  \bibfield  {author} {\bibinfo {author} {\bibfnamefont {S.~S.}\ \bibnamefont
  {Krishtopenko}}\ and\ \bibinfo {author} {\bibfnamefont {F.}~\bibnamefont
  {Teppe}},\ }\href {\doibase 10.1103/PhysRevB.105.125203} {\bibfield
  {journal} {\bibinfo  {journal} {Phys. Rev. B}\ }\textbf {\bibinfo {volume}
  {105}},\ \bibinfo {pages} {125203} (\bibinfo {year} {2022})}\BibitemShut
  {NoStop}%
\bibitem [{\citenamefont {Malcolm}\ and\ \citenamefont {Nicol}(2015)}]{Int4}%
  \BibitemOpen
  \bibfield  {author} {\bibinfo {author} {\bibfnamefont {J.~D.}\ \bibnamefont
  {Malcolm}}\ and\ \bibinfo {author} {\bibfnamefont {E.~J.}\ \bibnamefont
  {Nicol}},\ }\href {\doibase 10.1103/PhysRevB.92.035118} {\bibfield  {journal}
  {\bibinfo  {journal} {Phys. Rev. B}\ }\textbf {\bibinfo {volume} {92}},\
  \bibinfo {pages} {035118} (\bibinfo {year} {2015})}\BibitemShut {NoStop}%
\bibitem [{\citenamefont {Krishtopenko}\ \emph
  {et~al.}(2020{\natexlab{a}})\citenamefont {Krishtopenko}, \citenamefont
  {Antezza},\ and\ \citenamefont {Teppe}}]{Int5}%
  \BibitemOpen
  \bibfield  {author} {\bibinfo {author} {\bibfnamefont {S.~S.}\ \bibnamefont
  {Krishtopenko}}, \bibinfo {author} {\bibfnamefont {M.}~\bibnamefont
  {Antezza}}, \ and\ \bibinfo {author} {\bibfnamefont {F.}~\bibnamefont
  {Teppe}},\ }\href {\doibase 10.1088/1361-648x/ab6741} {\bibfield  {journal}
  {\bibinfo  {journal} {J. Phys.: Condens. Matter}\ }\textbf {\bibinfo {volume}
  {32}},\ \bibinfo {pages} {165501} (\bibinfo {year}
  {2020}{\natexlab{a}})}\BibitemShut {NoStop}%
\bibitem [{\citenamefont {Latussek}\ \emph {et~al.}(2005)\citenamefont
  {Latussek}, \citenamefont {Becker}, \citenamefont {Landwehr}, \citenamefont
  {Bini},\ and\ \citenamefont {Ulivi}}]{Int6}%
  \BibitemOpen
  \bibfield  {author} {\bibinfo {author} {\bibfnamefont {V.}~\bibnamefont
  {Latussek}}, \bibinfo {author} {\bibfnamefont {C.~R.}\ \bibnamefont
  {Becker}}, \bibinfo {author} {\bibfnamefont {G.}~\bibnamefont {Landwehr}},
  \bibinfo {author} {\bibfnamefont {R.}~\bibnamefont {Bini}}, \ and\ \bibinfo
  {author} {\bibfnamefont {L.}~\bibnamefont {Ulivi}},\ }\href {\doibase
  10.1103/PhysRevB.71.125305} {\bibfield  {journal} {\bibinfo  {journal} {Phys.
  Rev. B}\ }\textbf {\bibinfo {volume} {71}},\ \bibinfo {pages} {125305}
  (\bibinfo {year} {2005})}\BibitemShut {NoStop}%
\bibitem [{\citenamefont {Krishtopenko}\ \emph {et~al.}(2016)\citenamefont
  {Krishtopenko}, \citenamefont {Yahniuk}, \citenamefont {But}, \citenamefont
  {Gavrilenko}, \citenamefont {Knap},\ and\ \citenamefont {Teppe}}]{th1}%
  \BibitemOpen
  \bibfield  {author} {\bibinfo {author} {\bibfnamefont {S.~S.}\ \bibnamefont
  {Krishtopenko}}, \bibinfo {author} {\bibfnamefont {I.}~\bibnamefont
  {Yahniuk}}, \bibinfo {author} {\bibfnamefont {D.~B.}\ \bibnamefont {But}},
  \bibinfo {author} {\bibfnamefont {V.~I.}\ \bibnamefont {Gavrilenko}},
  \bibinfo {author} {\bibfnamefont {W.}~\bibnamefont {Knap}}, \ and\ \bibinfo
  {author} {\bibfnamefont {F.}~\bibnamefont {Teppe}},\ }\href {\doibase
  10.1103/PhysRevB.94.245402} {\bibfield  {journal} {\bibinfo  {journal} {Phys.
  Rev. B}\ }\textbf {\bibinfo {volume} {94}},\ \bibinfo {pages} {245402}
  (\bibinfo {year} {2016})}\BibitemShut {NoStop}%
\bibitem [{\citenamefont {Fu}\ and\ \citenamefont {Kane}(2007)}]{Int7}%
  \BibitemOpen
  \bibfield  {author} {\bibinfo {author} {\bibfnamefont {L.}~\bibnamefont
  {Fu}}\ and\ \bibinfo {author} {\bibfnamefont {C.~L.}\ \bibnamefont {Kane}},\
  }\href {\doibase 10.1103/PhysRevB.76.045302} {\bibfield  {journal} {\bibinfo
  {journal} {Phys. Rev. B}\ }\textbf {\bibinfo {volume} {76}},\ \bibinfo
  {pages} {045302} (\bibinfo {year} {2007})}\BibitemShut {NoStop}%
\bibitem [{\citenamefont {Virot}\ \emph {et~al.}(2013)\citenamefont {Virot},
  \citenamefont {Hayn}, \citenamefont {Richter},\ and\ \citenamefont {van~den
  Brink}}]{Int8}%
  \BibitemOpen
  \bibfield  {author} {\bibinfo {author} {\bibfnamefont {F.~m.~c.}\
  \bibnamefont {Virot}}, \bibinfo {author} {\bibfnamefont {R.}~\bibnamefont
  {Hayn}}, \bibinfo {author} {\bibfnamefont {M.}~\bibnamefont {Richter}}, \
  and\ \bibinfo {author} {\bibfnamefont {J.}~\bibnamefont {van~den Brink}},\
  }\href {\doibase 10.1103/PhysRevLett.111.146803} {\bibfield  {journal}
  {\bibinfo  {journal} {Phys. Rev. Lett.}\ }\textbf {\bibinfo {volume} {111}},\
  \bibinfo {pages} {146803} (\bibinfo {year} {2013})}\BibitemShut {NoStop}%
\bibitem [{\citenamefont {Li}\ \emph {et~al.}(2009)\citenamefont {Li},
  \citenamefont {Chu}, \citenamefont {Jain},\ and\ \citenamefont
  {Shen}}]{Int9}%
  \BibitemOpen
  \bibfield  {author} {\bibinfo {author} {\bibfnamefont {J.}~\bibnamefont
  {Li}}, \bibinfo {author} {\bibfnamefont {R.-L.}\ \bibnamefont {Chu}},
  \bibinfo {author} {\bibfnamefont {J.~K.}\ \bibnamefont {Jain}}, \ and\
  \bibinfo {author} {\bibfnamefont {S.-Q.}\ \bibnamefont {Shen}},\ }\href
  {\doibase 10.1103/PhysRevLett.102.136806} {\bibfield  {journal} {\bibinfo
  {journal} {Phys. Rev. Lett.}\ }\textbf {\bibinfo {volume} {102}},\ \bibinfo
  {pages} {136806} (\bibinfo {year} {2009})}\BibitemShut {NoStop}%
\bibitem [{\citenamefont {Groth}\ \emph {et~al.}(2009)\citenamefont {Groth},
  \citenamefont {Wimmer}, \citenamefont {Akhmerov}, \citenamefont
  {Tworzyd\l{}o},\ and\ \citenamefont {Beenakker}}]{Int10}%
  \BibitemOpen
  \bibfield  {author} {\bibinfo {author} {\bibfnamefont {C.~W.}\ \bibnamefont
  {Groth}}, \bibinfo {author} {\bibfnamefont {M.}~\bibnamefont {Wimmer}},
  \bibinfo {author} {\bibfnamefont {A.~R.}\ \bibnamefont {Akhmerov}}, \bibinfo
  {author} {\bibfnamefont {J.}~\bibnamefont {Tworzyd\l{}o}}, \ and\ \bibinfo
  {author} {\bibfnamefont {C.~W.~J.}\ \bibnamefont {Beenakker}},\ }\href
  {\doibase 10.1103/PhysRevLett.103.196805} {\bibfield  {journal} {\bibinfo
  {journal} {Phys. Rev. Lett.}\ }\textbf {\bibinfo {volume} {103}},\ \bibinfo
  {pages} {196805} (\bibinfo {year} {2009})}\BibitemShut {NoStop}%
\bibitem [{\citenamefont {Bernevig}\ \emph {et~al.}(2006)\citenamefont
  {Bernevig}, \citenamefont {Hughes},\ and\ \citenamefont {Zhang}}]{Int11}%
  \BibitemOpen
  \bibfield  {author} {\bibinfo {author} {\bibfnamefont {B.~A.}\ \bibnamefont
  {Bernevig}}, \bibinfo {author} {\bibfnamefont {T.~L.}\ \bibnamefont
  {Hughes}}, \ and\ \bibinfo {author} {\bibfnamefont {S.-C.}\ \bibnamefont
  {Zhang}},\ }\href {\doibase 10.1126/science.1133734} {\bibfield  {journal}
  {\bibinfo  {journal} {Science}\ }\textbf {\bibinfo {volume} {314}},\ \bibinfo
  {pages} {1757} (\bibinfo {year} {2006})}\BibitemShut {NoStop}%
\bibitem [{\citenamefont {Shen}\ \emph {et~al.}(2011)\citenamefont {Shen},
  \citenamefont {Shan},\ and\ \citenamefont {Lu}}]{Int12}%
  \BibitemOpen
  \bibfield  {author} {\bibinfo {author} {\bibfnamefont {S.-Q.}\ \bibnamefont
  {Shen}}, \bibinfo {author} {\bibfnamefont {W.-Y.}\ \bibnamefont {Shan}}, \
  and\ \bibinfo {author} {\bibfnamefont {H.-Z.}\ \bibnamefont {Lu}},\ }\href
  {\doibase 10.1142/S2010324711000057} {\bibfield  {journal} {\bibinfo
  {journal} {SPIN}\ }\textbf {\bibinfo {volume} {01}},\ \bibinfo {pages} {33}
  (\bibinfo {year} {2011})}\BibitemShut {NoStop}%
\bibitem [{\citenamefont {Guo}\ \emph {et~al.}(2010)\citenamefont {Guo},
  \citenamefont {Rosenberg}, \citenamefont {Refael},\ and\ \citenamefont
  {Franz}}]{TAI3Da}%
  \BibitemOpen
  \bibfield  {author} {\bibinfo {author} {\bibfnamefont {H.-M.}\ \bibnamefont
  {Guo}}, \bibinfo {author} {\bibfnamefont {G.}~\bibnamefont {Rosenberg}},
  \bibinfo {author} {\bibfnamefont {G.}~\bibnamefont {Refael}}, \ and\ \bibinfo
  {author} {\bibfnamefont {M.}~\bibnamefont {Franz}},\ }\href {\doibase
  10.1103/PhysRevLett.105.216601} {\bibfield  {journal} {\bibinfo  {journal}
  {Phys. Rev. Lett.}\ }\textbf {\bibinfo {volume} {105}},\ \bibinfo {pages}
  {216601} (\bibinfo {year} {2010})}\BibitemShut {NoStop}%
\bibitem [{\citenamefont {Kobayashi}\ \emph {et~al.}(2013)\citenamefont
  {Kobayashi}, \citenamefont {Ohtsuki},\ and\ \citenamefont {Imura}}]{TAI3Db}%
  \BibitemOpen
  \bibfield  {author} {\bibinfo {author} {\bibfnamefont {K.}~\bibnamefont
  {Kobayashi}}, \bibinfo {author} {\bibfnamefont {T.}~\bibnamefont {Ohtsuki}},
  \ and\ \bibinfo {author} {\bibfnamefont {K.-I.}\ \bibnamefont {Imura}},\
  }\href {\doibase 10.1103/PhysRevLett.110.236803} {\bibfield  {journal}
  {\bibinfo  {journal} {Phys. Rev. Lett.}\ }\textbf {\bibinfo {volume} {110}},\
  \bibinfo {pages} {236803} (\bibinfo {year} {2013})}\BibitemShut {NoStop}%
\bibitem [{\citenamefont {Ryu}\ and\ \citenamefont {Nomura}(2012)}]{TAI3Dc}%
  \BibitemOpen
  \bibfield  {author} {\bibinfo {author} {\bibfnamefont {S.}~\bibnamefont
  {Ryu}}\ and\ \bibinfo {author} {\bibfnamefont {K.}~\bibnamefont {Nomura}},\
  }\href {\doibase 10.1103/PhysRevB.85.155138} {\bibfield  {journal} {\bibinfo
  {journal} {Phys. Rev. B}\ }\textbf {\bibinfo {volume} {85}},\ \bibinfo
  {pages} {155138} (\bibinfo {year} {2012})}\BibitemShut {NoStop}%
\bibitem [{\citenamefont {Chen}\ \emph {et~al.}(2015)\citenamefont {Chen},
  \citenamefont {Song}, \citenamefont {Jiang}, \citenamefont {Sun},
  \citenamefont {Wang},\ and\ \citenamefont {Xie}}]{TAI3Dd}%
  \BibitemOpen
  \bibfield  {author} {\bibinfo {author} {\bibfnamefont {C.-Z.}\ \bibnamefont
  {Chen}}, \bibinfo {author} {\bibfnamefont {J.}~\bibnamefont {Song}}, \bibinfo
  {author} {\bibfnamefont {H.}~\bibnamefont {Jiang}}, \bibinfo {author}
  {\bibfnamefont {Q.-f.}\ \bibnamefont {Sun}}, \bibinfo {author} {\bibfnamefont
  {Z.}~\bibnamefont {Wang}}, \ and\ \bibinfo {author} {\bibfnamefont {X.~C.}\
  \bibnamefont {Xie}},\ }\href {\doibase 10.1103/PhysRevLett.115.246603}
  {\bibfield  {journal} {\bibinfo  {journal} {Phys. Rev. Lett.}\ }\textbf
  {\bibinfo {volume} {115}},\ \bibinfo {pages} {246603} (\bibinfo {year}
  {2015})}\BibitemShut {NoStop}%
\bibitem [{\citenamefont {Shapourian}\ and\ \citenamefont
  {Hughes}(2016)}]{TAI3De}%
  \BibitemOpen
  \bibfield  {author} {\bibinfo {author} {\bibfnamefont {H.}~\bibnamefont
  {Shapourian}}\ and\ \bibinfo {author} {\bibfnamefont {T.~L.}\ \bibnamefont
  {Hughes}},\ }\href {\doibase 10.1103/PhysRevB.93.075108} {\bibfield
  {journal} {\bibinfo  {journal} {Phys. Rev. B}\ }\textbf {\bibinfo {volume}
  {93}},\ \bibinfo {pages} {075108} (\bibinfo {year} {2016})}\BibitemShut
  {NoStop}%
\bibitem [{\citenamefont {Bera}\ \emph {et~al.}(2016)\citenamefont {Bera},
  \citenamefont {Sau},\ and\ \citenamefont {Roy}}]{TAI3Df}%
  \BibitemOpen
  \bibfield  {author} {\bibinfo {author} {\bibfnamefont {S.}~\bibnamefont
  {Bera}}, \bibinfo {author} {\bibfnamefont {J.~D.}\ \bibnamefont {Sau}}, \
  and\ \bibinfo {author} {\bibfnamefont {B.}~\bibnamefont {Roy}},\ }\href
  {\doibase 10.1103/PhysRevB.93.201302} {\bibfield  {journal} {\bibinfo
  {journal} {Phys. Rev. B}\ }\textbf {\bibinfo {volume} {93}},\ \bibinfo
  {pages} {201302} (\bibinfo {year} {2016})}\BibitemShut {NoStop}%
\bibitem [{\citenamefont {Park}\ \emph {et~al.}(2017)\citenamefont {Park},
  \citenamefont {Basa},\ and\ \citenamefont {Gilbert}}]{TAI3Dg}%
  \BibitemOpen
  \bibfield  {author} {\bibinfo {author} {\bibfnamefont {M.~J.}\ \bibnamefont
  {Park}}, \bibinfo {author} {\bibfnamefont {B.}~\bibnamefont {Basa}}, \ and\
  \bibinfo {author} {\bibfnamefont {M.~J.}\ \bibnamefont {Gilbert}},\ }\href
  {\doibase 10.1103/PhysRevB.95.094201} {\bibfield  {journal} {\bibinfo
  {journal} {Phys. Rev. B}\ }\textbf {\bibinfo {volume} {95}},\ \bibinfo
  {pages} {094201} (\bibinfo {year} {2017})}\BibitemShut {NoStop}%
\bibitem [{TAI()}]{TAI3Dh}%
  \BibitemOpen
  \href@noop {} {\ }\BibitemShut {NoStop}%
\bibitem [{\citenamefont {Chen}\ \emph {et~al.}(2012)\citenamefont {Chen},
  \citenamefont {Liu}, \citenamefont {Lin}, \citenamefont {Zhang},\ and\
  \citenamefont {Jiang}}]{Int13}%
  \BibitemOpen
  \bibfield  {author} {\bibinfo {author} {\bibfnamefont {L.}~\bibnamefont
  {Chen}}, \bibinfo {author} {\bibfnamefont {Q.}~\bibnamefont {Liu}}, \bibinfo
  {author} {\bibfnamefont {X.}~\bibnamefont {Lin}}, \bibinfo {author}
  {\bibfnamefont {X.}~\bibnamefont {Zhang}}, \ and\ \bibinfo {author}
  {\bibfnamefont {X.}~\bibnamefont {Jiang}},\ }\href {\doibase
  10.1088/1367-2630/14/4/043028} {\bibfield  {journal} {\bibinfo  {journal}
  {New J. of Phys.}\ }\textbf {\bibinfo {volume} {14}},\ \bibinfo {pages}
  {043028} (\bibinfo {year} {2012})}\BibitemShut {NoStop}%
\bibitem [{\citenamefont {Girschik}\ \emph {et~al.}(2013)\citenamefont
  {Girschik}, \citenamefont {Libisch},\ and\ \citenamefont {Rotter}}]{Int14}%
  \BibitemOpen
  \bibfield  {author} {\bibinfo {author} {\bibfnamefont {A.}~\bibnamefont
  {Girschik}}, \bibinfo {author} {\bibfnamefont {F.}~\bibnamefont {Libisch}}, \
  and\ \bibinfo {author} {\bibfnamefont {S.}~\bibnamefont {Rotter}},\ }\href
  {\doibase 10.1103/PhysRevB.88.014201} {\bibfield  {journal} {\bibinfo
  {journal} {Phys. Rev. B}\ }\textbf {\bibinfo {volume} {88}},\ \bibinfo
  {pages} {014201} (\bibinfo {year} {2013})}\BibitemShut {NoStop}%
\bibitem [{\citenamefont {Krishtopenko}\ \emph
  {et~al.}(2020{\natexlab{b}})\citenamefont {Krishtopenko}, \citenamefont
  {Antezza},\ and\ \citenamefont {Teppe}}]{Int15}%
  \BibitemOpen
  \bibfield  {author} {\bibinfo {author} {\bibfnamefont {S.~S.}\ \bibnamefont
  {Krishtopenko}}, \bibinfo {author} {\bibfnamefont {M.}~\bibnamefont
  {Antezza}}, \ and\ \bibinfo {author} {\bibfnamefont {F.}~\bibnamefont
  {Teppe}},\ }\href {\doibase 10.1103/PhysRevB.101.205424} {\bibfield
  {journal} {\bibinfo  {journal} {Phys. Rev. B}\ }\textbf {\bibinfo {volume}
  {101}},\ \bibinfo {pages} {205424} (\bibinfo {year}
  {2020}{\natexlab{b}})}\BibitemShut {NoStop}%
\bibitem [{\citenamefont {Rogalski}(2005)}]{Int22}%
  \BibitemOpen
  \bibfield  {author} {\bibinfo {author} {\bibfnamefont {A.}~\bibnamefont
  {Rogalski}},\ }\href {\doibase 10.1088/0034-4885/68/10/r01} {\bibfield
  {journal} {\bibinfo  {journal} {Rep. Prog. Phys.}\ }\textbf {\bibinfo
  {volume} {68}},\ \bibinfo {pages} {2267} (\bibinfo {year}
  {2005})}\BibitemShut {NoStop}%
\bibitem [{\citenamefont {Ivanov-Omskii}\ \emph {et~al.}(2009)\citenamefont
  {Ivanov-Omskii}, \citenamefont {Bazhenov}, \citenamefont {Mynbaev},
  \citenamefont {Smirnov}, \citenamefont {Varavin}, \citenamefont {Mikhailov},\
  and\ \citenamefont {Sidorov}}]{Int23}%
  \BibitemOpen
  \bibfield  {author} {\bibinfo {author} {\bibfnamefont {V.}~\bibnamefont
  {Ivanov-Omskii}}, \bibinfo {author} {\bibfnamefont {N.}~\bibnamefont
  {Bazhenov}}, \bibinfo {author} {\bibfnamefont {K.}~\bibnamefont {Mynbaev}},
  \bibinfo {author} {\bibfnamefont {V.}~\bibnamefont {Smirnov}}, \bibinfo
  {author} {\bibfnamefont {V.}~\bibnamefont {Varavin}}, \bibinfo {author}
  {\bibfnamefont {N.}~\bibnamefont {Mikhailov}}, \ and\ \bibinfo {author}
  {\bibfnamefont {G.}~\bibnamefont {Sidorov}},\ }\href {\doibase
  https://doi.org/10.1016/j.physb.2009.08.210} {\bibfield  {journal} {\bibinfo
  {journal} {Phys. B: Condens. Matter}\ }\textbf {\bibinfo {volume} {404}},\
  \bibinfo {pages} {5035} (\bibinfo {year} {2009})}\BibitemShut {NoStop}%
\bibitem [{\citenamefont {Belenky}\ \emph {et~al.}(2011)\citenamefont
  {Belenky}, \citenamefont {Donetsky}, \citenamefont {Kipshidze}, \citenamefont
  {Wang}, \citenamefont {Shterengas}, \citenamefont {Sarney},\ and\
  \citenamefont {Svensson}}]{Int24}%
  \BibitemOpen
  \bibfield  {author} {\bibinfo {author} {\bibfnamefont {G.}~\bibnamefont
  {Belenky}}, \bibinfo {author} {\bibfnamefont {D.}~\bibnamefont {Donetsky}},
  \bibinfo {author} {\bibfnamefont {G.}~\bibnamefont {Kipshidze}}, \bibinfo
  {author} {\bibfnamefont {D.}~\bibnamefont {Wang}}, \bibinfo {author}
  {\bibfnamefont {L.}~\bibnamefont {Shterengas}}, \bibinfo {author}
  {\bibfnamefont {W.~L.}\ \bibnamefont {Sarney}}, \ and\ \bibinfo {author}
  {\bibfnamefont {S.~P.}\ \bibnamefont {Svensson}},\ }\href {\doibase
  10.1063/1.3650473} {\bibfield  {journal} {\bibinfo  {journal} {Appl. Phys.
  Lett.}\ }\textbf {\bibinfo {volume} {99}},\ \bibinfo {pages} {141116}
  (\bibinfo {year} {2011})}\BibitemShut {NoStop}%
\bibitem [{\citenamefont {Svensson}\ \emph {et~al.}(2012)\citenamefont
  {Svensson}, \citenamefont {Sarney}, \citenamefont {Hier}, \citenamefont
  {Lin}, \citenamefont {Wang}, \citenamefont {Donetsky}, \citenamefont
  {Shterengas}, \citenamefont {Kipshidze},\ and\ \citenamefont
  {Belenky}}]{Int25}%
  \BibitemOpen
  \bibfield  {author} {\bibinfo {author} {\bibfnamefont {S.~P.}\ \bibnamefont
  {Svensson}}, \bibinfo {author} {\bibfnamefont {W.~L.}\ \bibnamefont
  {Sarney}}, \bibinfo {author} {\bibfnamefont {H.}~\bibnamefont {Hier}},
  \bibinfo {author} {\bibfnamefont {Y.}~\bibnamefont {Lin}}, \bibinfo {author}
  {\bibfnamefont {D.}~\bibnamefont {Wang}}, \bibinfo {author} {\bibfnamefont
  {D.}~\bibnamefont {Donetsky}}, \bibinfo {author} {\bibfnamefont
  {L.}~\bibnamefont {Shterengas}}, \bibinfo {author} {\bibfnamefont
  {G.}~\bibnamefont {Kipshidze}}, \ and\ \bibinfo {author} {\bibfnamefont
  {G.}~\bibnamefont {Belenky}},\ }\href {\doibase 10.1103/PhysRevB.86.245205}
  {\bibfield  {journal} {\bibinfo  {journal} {Phys. Rev. B}\ }\textbf {\bibinfo
  {volume} {86}},\ \bibinfo {pages} {245205} (\bibinfo {year}
  {2012})}\BibitemShut {NoStop}%
\bibitem [{\citenamefont {Suchalkin}\ \emph {et~al.}(2016)\citenamefont
  {Suchalkin}, \citenamefont {Ludwig}, \citenamefont {Belenky}, \citenamefont
  {Laikhtman}, \citenamefont {Kipshidze}, \citenamefont {Lin}, \citenamefont
  {Shterengas}, \citenamefont {Smirnov}, \citenamefont {Luryi}, \citenamefont
  {Sarney},\ and\ \citenamefont {Svensson}}]{Int26}%
  \BibitemOpen
  \bibfield  {author} {\bibinfo {author} {\bibfnamefont {S.}~\bibnamefont
  {Suchalkin}}, \bibinfo {author} {\bibfnamefont {J.}~\bibnamefont {Ludwig}},
  \bibinfo {author} {\bibfnamefont {G.}~\bibnamefont {Belenky}}, \bibinfo
  {author} {\bibfnamefont {B.}~\bibnamefont {Laikhtman}}, \bibinfo {author}
  {\bibfnamefont {G.}~\bibnamefont {Kipshidze}}, \bibinfo {author}
  {\bibfnamefont {Y.}~\bibnamefont {Lin}}, \bibinfo {author} {\bibfnamefont
  {L.}~\bibnamefont {Shterengas}}, \bibinfo {author} {\bibfnamefont
  {D.}~\bibnamefont {Smirnov}}, \bibinfo {author} {\bibfnamefont
  {S.}~\bibnamefont {Luryi}}, \bibinfo {author} {\bibfnamefont {W.~L.}\
  \bibnamefont {Sarney}}, \ and\ \bibinfo {author} {\bibfnamefont {S.~P.}\
  \bibnamefont {Svensson}},\ }\href {\doibase 10.1088/0022-3727/49/10/105101}
  {\bibfield  {journal} {\bibinfo  {journal} {J. Phys. D: Appl. Phys.}\
  }\textbf {\bibinfo {volume} {49}},\ \bibinfo {pages} {105101} (\bibinfo
  {year} {2016})}\BibitemShut {NoStop}%
\bibitem [{\citenamefont {Luttinger}\ and\ \citenamefont {Kohn}(1955)}]{th2}%
  \BibitemOpen
  \bibfield  {author} {\bibinfo {author} {\bibfnamefont {J.~M.}\ \bibnamefont
  {Luttinger}}\ and\ \bibinfo {author} {\bibfnamefont {W.}~\bibnamefont
  {Kohn}},\ }\href {\doibase 10.1103/PhysRev.97.869} {\bibfield  {journal}
  {\bibinfo  {journal} {Phys. Rev.}\ }\textbf {\bibinfo {volume} {97}},\
  \bibinfo {pages} {869} (\bibinfo {year} {1955})}\BibitemShut {NoStop}%
\bibitem [{\citenamefont {Dresselhaus}(1955)}]{th3}%
  \BibitemOpen
  \bibfield  {author} {\bibinfo {author} {\bibfnamefont {G.}~\bibnamefont
  {Dresselhaus}},\ }\href {\doibase 10.1103/PhysRev.100.580} {\bibfield
  {journal} {\bibinfo  {journal} {Phys. Rev.}\ }\textbf {\bibinfo {volume}
  {100}},\ \bibinfo {pages} {580} (\bibinfo {year} {1955})}\BibitemShut
  {NoStop}%
\bibitem [{\citenamefont {Zimmermann}\ and\ \citenamefont
  {Schindler}(2009)}]{Int16}%
  \BibitemOpen
  \bibfield  {author} {\bibinfo {author} {\bibfnamefont {R.}~\bibnamefont
  {Zimmermann}}\ and\ \bibinfo {author} {\bibfnamefont {C.}~\bibnamefont
  {Schindler}},\ }\href {\doibase 10.1103/PhysRevB.80.144202} {\bibfield
  {journal} {\bibinfo  {journal} {Phys. Rev. B}\ }\textbf {\bibinfo {volume}
  {80}},\ \bibinfo {pages} {144202} (\bibinfo {year} {2009})}\BibitemShut
  {NoStop}%
\bibitem [{\citenamefont {Tkachov}\ and\ \citenamefont
  {Hankiewicz}(2011)}]{Int17}%
  \BibitemOpen
  \bibfield  {author} {\bibinfo {author} {\bibfnamefont {G.}~\bibnamefont
  {Tkachov}}\ and\ \bibinfo {author} {\bibfnamefont {E.~M.}\ \bibnamefont
  {Hankiewicz}},\ }\href {\doibase 10.1103/PhysRevB.84.035444} {\bibfield
  {journal} {\bibinfo  {journal} {Phys. Rev. B}\ }\textbf {\bibinfo {volume}
  {84}},\ \bibinfo {pages} {035444} (\bibinfo {year} {2011})}\BibitemShut
  {NoStop}%
\bibitem [{\citenamefont {Song}\ \emph {et~al.}(2012)\citenamefont {Song},
  \citenamefont {Liu}, \citenamefont {Jiang}, \citenamefont {Sun},\ and\
  \citenamefont {Xie}}]{Int18}%
  \BibitemOpen
  \bibfield  {author} {\bibinfo {author} {\bibfnamefont {J.}~\bibnamefont
  {Song}}, \bibinfo {author} {\bibfnamefont {H.}~\bibnamefont {Liu}}, \bibinfo
  {author} {\bibfnamefont {H.}~\bibnamefont {Jiang}}, \bibinfo {author}
  {\bibfnamefont {Q.-f.}\ \bibnamefont {Sun}}, \ and\ \bibinfo {author}
  {\bibfnamefont {X.~C.}\ \bibnamefont {Xie}},\ }\href {\doibase
  10.1103/PhysRevB.85.195125} {\bibfield  {journal} {\bibinfo  {journal} {Phys.
  Rev. B}\ }\textbf {\bibinfo {volume} {85}},\ \bibinfo {pages} {195125}
  (\bibinfo {year} {2012})}\BibitemShut {NoStop}%
\bibitem [{\citenamefont {Arimura}\ and\ \citenamefont {Ando}(2012)}]{Int19}%
  \BibitemOpen
  \bibfield  {author} {\bibinfo {author} {\bibfnamefont {Y.}~\bibnamefont
  {Arimura}}\ and\ \bibinfo {author} {\bibfnamefont {T.}~\bibnamefont {Ando}},\
  }\href {\doibase 10.1143/JPSJ.81.024702} {\bibfield  {journal} {\bibinfo
  {journal} {J. Phys. Soc. Jpn.}\ }\textbf {\bibinfo {volume} {81}},\ \bibinfo
  {pages} {024702} (\bibinfo {year} {2012})}\BibitemShut {NoStop}%
\bibitem [{\citenamefont {Ando}(2015)}]{Int20}%
  \BibitemOpen
  \bibfield  {author} {\bibinfo {author} {\bibfnamefont {T.}~\bibnamefont
  {Ando}},\ }\href {\doibase 10.7566/JPSJ.84.114705} {\bibfield  {journal}
  {\bibinfo  {journal} {J. Phys. Soc. Jpn.}\ }\textbf {\bibinfo {volume}
  {84}},\ \bibinfo {pages} {114705} (\bibinfo {year} {2015})}\BibitemShut
  {NoStop}%
\bibitem [{\citenamefont {Rostami}\ and\ \citenamefont
  {Cappelluti}(2017)}]{Int21}%
  \BibitemOpen
  \bibfield  {author} {\bibinfo {author} {\bibfnamefont {H.}~\bibnamefont
  {Rostami}}\ and\ \bibinfo {author} {\bibfnamefont {E.}~\bibnamefont
  {Cappelluti}},\ }\href {\doibase 10.1103/PhysRevB.96.054205} {\bibfield
  {journal} {\bibinfo  {journal} {Phys. Rev. B}\ }\textbf {\bibinfo {volume}
  {96}},\ \bibinfo {pages} {054205} (\bibinfo {year} {2017})}\BibitemShut
  {NoStop}%
\bibitem [{\citenamefont {Zhang}\ \emph {et~al.}(2009)\citenamefont {Zhang},
  \citenamefont {Liu}, \citenamefont {Qi}, \citenamefont {Dai}, \citenamefont
  {Fang},\ and\ \citenamefont {Zhang}}]{th4}%
  \BibitemOpen
  \bibfield  {author} {\bibinfo {author} {\bibfnamefont {H.}~\bibnamefont
  {Zhang}}, \bibinfo {author} {\bibfnamefont {C.-X.}\ \bibnamefont {Liu}},
  \bibinfo {author} {\bibfnamefont {X.-L.}\ \bibnamefont {Qi}}, \bibinfo
  {author} {\bibfnamefont {X.}~\bibnamefont {Dai}}, \bibinfo {author}
  {\bibfnamefont {Z.}~\bibnamefont {Fang}}, \ and\ \bibinfo {author}
  {\bibfnamefont {S.-C.}\ \bibnamefont {Zhang}},\ }\href {\doibase
  10.1038/nphys1270} {\bibfield  {journal} {\bibinfo  {journal} {Nature Phys.}\
  }\textbf {\bibinfo {volume} {5}},\ \bibinfo {pages} {438} (\bibinfo {year}
  {2009})}\BibitemShut {NoStop}%
\bibitem [{\citenamefont {Shan}\ \emph {et~al.}(2010)\citenamefont {Shan},
  \citenamefont {Lu},\ and\ \citenamefont {Shen}}]{th5}%
  \BibitemOpen
  \bibfield  {author} {\bibinfo {author} {\bibfnamefont {W.-Y.}\ \bibnamefont
  {Shan}}, \bibinfo {author} {\bibfnamefont {H.-Z.}\ \bibnamefont {Lu}}, \ and\
  \bibinfo {author} {\bibfnamefont {S.-Q.}\ \bibnamefont {Shen}},\ }\href
  {\doibase 10.1088/1367-2630/12/4/043048} {\bibfield  {journal} {\bibinfo
  {journal} {New J. Phys.}\ }\textbf {\bibinfo {volume} {12}},\ \bibinfo
  {pages} {043048} (\bibinfo {year} {2010})}\BibitemShut {NoStop}%
\bibitem [{\citenamefont {Lu}\ \emph {et~al.}(2010)\citenamefont {Lu},
  \citenamefont {Shan}, \citenamefont {Yao}, \citenamefont {Niu},\ and\
  \citenamefont {Shen}}]{th6}%
  \BibitemOpen
  \bibfield  {author} {\bibinfo {author} {\bibfnamefont {H.-Z.}\ \bibnamefont
  {Lu}}, \bibinfo {author} {\bibfnamefont {W.-Y.}\ \bibnamefont {Shan}},
  \bibinfo {author} {\bibfnamefont {W.}~\bibnamefont {Yao}}, \bibinfo {author}
  {\bibfnamefont {Q.}~\bibnamefont {Niu}}, \ and\ \bibinfo {author}
  {\bibfnamefont {S.-Q.}\ \bibnamefont {Shen}},\ }\href {\doibase
  10.1103/PhysRevB.81.115407} {\bibfield  {journal} {\bibinfo  {journal} {Phys.
  Rev. B}\ }\textbf {\bibinfo {volume} {81}},\ \bibinfo {pages} {115407}
  (\bibinfo {year} {2010})}\BibitemShut {NoStop}%
\end{thebibliography}
%

\end{document}